\documentclass[fleqn,usenatbib,useAMS]{mnras}

\usepackage{graphicx}	% Including figure files
\usepackage{amsmath}	% Advanced maths commands
\usepackage{amssymb}	% Extra maths symbols
\usepackage{multicol}        % Multi-column entries in tables
\usepackage{bm}		% Bold maths symbols, including upright Greek
\usepackage{pdflscape}	% Landscape pages
\usepackage{threeparttable}

\usepackage[T1]{fontenc}
\usepackage{ae,aecompl}

\usepackage{newtxtext,newtxmath}
\title[PHANGS--HST Multi-Scale Stellar Associations]{Multi-Scale Stellar Associations across the Star Formation Hierarchy in PHANGS--HST Nearby Galaxies: Methodology and Properties}

\author[Larson et al.]
{Kirsten~L.~Larson,$^{1,2}$\thanks{Contact e-mail: \href{mailto:kilarson@stsci.edu}{kilarson@stsci.edu}}
Janice~C.~Lee,$^{2,3}$
David~A.~Thilker,$^{4}$
Bradley~C.~Whitmore,$^{5}$
\newauthor
Sinan~Deger,$^{2,6}$
James~Lilly,$^{7}$
Rupali~Chandar,$^{8}$
Daniel~A.~Dale,$^{7}$
Frank~Bigiel,$^{9}$
\newauthor
Kathryn~Grasha,$^{10}$
Brent~Groves,$^{11,10}$ 
Ralf~S.~Klessen,$^{12,13}$
Kathryn~Kreckel,$^{14}$
\newauthor
J.~M.~Diederik~Kruijssen,$^{15}$
Adam~K.~Leroy,$^{16}$
Hsi-An~Pan,$^{17}$ %
Erik~Rosolowsky,$^{18}$
\newauthor
Eva Schinnerer,$^{17}$
Andreas~Schruba,$^{19}$
Elizabeth~J.~Watkins,$^{14}$
Thomas~G.~Williams,$^{20, 17}$ 
\\
$^{1}$AURA for the European Space Agency (ESA), Space Telescope Science Institute, 3700 San Martin Drive, Baltimore, MD, USA\\
$^{2}$Caltech/IPAC, California Institute of Technology, Pasadena, CA, USA\\
$^{3}$Gemini Observatory/NSF’s NOIRLab, 950 N. Cherry Avenue, Tucson, AZ, USA\\
$^{4}$Department of Physics and Astronomy, The Johns Hopkins University, Baltimore, MD, USA\\
$^{5}$Space Telescope Science Institute, 3700 San Martin Drive, Baltimore, MD, USA\\
$^{6}$The Oskar Klein Centre for Cosmoparticle Physics, Department of Physics, Stockholm University, AlbaNova, Stockholm, SE-106 91, Sweden\\
$^{7}$Department of Physics and Astronomy, University of Wyoming, Laramie, WY, USA\\
$^{8}$Department of Physics and Astronomy, University of Toledo, Toledo, OH USA\\
$^{9}$Argelander-Institut f\"ur Astronomie, Universit\"at Bonn, Auf dem H\"ugel 71, 53121 Bonn, Germany \\
$^{10}$Research School of Astronomy and Astrophysics, Australian National University, Weston Creek, ACT 2611, Australia\\
$^{11}$International Centre for Radio Astronomy Research, University of Western Australia, 35 Stirling Highway, Crawley, WA 6009, Australia\\
$^{12}$Universit\"{a}t Heidelberg, Zentrum f\"{u}r Astronomie, Institut f\"{u}r Theoretische Astrophysik, Albert-Ueberle-Straße 2, D-69120 Heidelberg, Germany \\
$^{13}$Universit\"{a}t Heidelberg, Interdisziplin\"{a}res Zentrum f\"{u}r Wissenschaftliches Rechnen, Im Neuenheimer Feld 205, D-69120 Heidelberg, Germany\\
$^{14}$Astronomisches Rechen-Institut, Zentrum f\"{u}r Astronomie der Universit\"{a}t Heidelberg, M\"{o}nchhofstra\ss e 12-14, 69120 Heidelberg, Germany\\
$^{15}$Cosmic Origins Of Life (COOL) Research DAO, coolresearch.io \\
$^{16}$Department of Astronomy, The Ohio State University, 140 West 18th Avenue, Columbus, Ohio 43210, USA\\
$^{17}$Max-Planck-Institut f\"{u}r Astronomie, K\"{o}nigstuhl 17, D-69117, Heidelberg, Germany\\
$^{18}$4-183 CCIS, University of Alberta, Edmonton, Alberta, Canada\\
$^{19}$Max-Planck-Institut f{\"{u}}r extraterrestrische Physik, Giessenbachstra{\ss}e~1, D-85748 Garching, Germany\\
$^{20}$Sub-department of Astrophysics, Department of Physics, University of Oxford, Keble Road, Oxford OX1 3RH, UK\\
}

\pubyear{2022}

\begin{document}
\label{firstpage}
\pagerange{\pageref{firstpage}--\pageref{lastpage}}
\maketitle

% Abstract of the paper

\begin{abstract} We develop a method to identify and determine the physical properties of stellar associations using \textit{Hubble Space Telescope} (\textit{HST}) \textit{NUV-U-B-V-I} imaging of nearby galaxies from the PHANGS--HST survey.  
We apply a watershed algorithm to density maps constructed from point source catalogues Gaussian smoothed to multiple physical scales from 8 to 64~pc. 
We develop our method on two galaxies that span the distance range in the PHANGS--HST sample: NGC~3351 (10~Mpc), NGC~1566 (18~Mpc). 
We test our algorithm with different parameters such as the choice of detection band for the point source catalogue (\textit{NUV} or~\textit{V}), source density image filtering methods, and absolute magnitude limits.  
We characterise the properties of the resulting multi-scale associations, including sizes, number of tracer stars, number of associations, photometry, as well as ages, masses, and reddening from Spectral Energy Distribution fitting.  Our method successfully identifies structures that occupy loci in the \textit{UBVI} colour--colour diagram consistent with previously published catalogues of clusters and associations. The median ages of the associations increases  from $\log (\mathrm{age/yr}) = 6.6$ to $\log (\mathrm{age/yr}) = 6.9$ as the spatial scale increases from 8~pc to 64~pc for both galaxies. We find that the youngest stellar associations, with ages $<3$~Myr, indeed closely trace \textsc{H\,ii} regions in H$\alpha$ imaging, and that older associations are increasingly anti-correlated with the H$\alpha$ emission.  Owing to our new method, the PHANGS--HST multi-scale associations provide a far more complete census of recent star formation activity than found with previous cluster and compact association catalogues. 
The method presented here will be applied to the full sample of 38 PHANGS--HST galaxies.% to produce multi-scale association catalogues to be publicly released together with the scripts used.
 
%We take a first look at the correlation of our watershed-based multi-scale associations with the rather differently identified LEGUS Class~3 objects (compact associations), PHANGS ground-based H$\alpha$ imaging, and ALMA CO cloud catalogues. Most of the LEGUS Class~3 objects are contained within the watershed defined associations by the \mbox{16-pc} scale level with 93\% for NGC~3351 and 68\% for NGC~1566. 
% The percentage of ALMA CO clouds that have overlap with the \mbox{64-pc} stellar associations, which is the scale level closest to the ALMA beam size, is $\sim$58\% for both galaxies, though with significant local environmental variation. 

\end{abstract}

% Select between one and six entries from the list of approved keywords.
% Don't make up new ones.

%%%%%%%%%%%%%%%%% BODY OF PAPER %%%%%%%%%%%%%%%%%%
\section{Introduction}

Star formation occurs over a large range in physical scales, from parsec-sized molecular clumps to large star-forming regions of a few 100~pcs in the disks and nuclear regions of galaxies. 
Resolving star-forming regions and linking emission from young stars to the physical conditions and gas supply is critical to understanding the ubiquity of the Kennicutt--Schmidt relation \citep{schmidt59,kennicutt98} and the variation of star formation efficiency across galaxies and galaxy types.
To fully understand the star formation process, we need to connect the spatial distribution and timescales of stars from their formation in molecular clouds to the destruction of their birth clouds and the removal of gas from feedback. 
The UV-optical imaging obtained for nearby galaxies by the PHANGS--HST programme \citep[][]{lee22} will allow us to track the timescales for star formation structure evolution from stars at parsec scales, to stellar populations on 100~pc scales, up to galactic scales.

The majority of star formation generally occurs in stellar associations \citep[][and references therein]{lada03,ward18,gouliermis18, ward20, wright20} with the compact star clusters forming at the densest peaks of this stellar hierarchy \citep{bruce08, kruijssen12}. Star clusters account for ${\sim}1{-}50$~per~cent of the star formation in galaxies \citep{kruijssen12, adamo15, cliff16, chandar17, krumholz19, adamo20, krause20}. In order to get a more complete census of star formation in galaxies, we must therefore trace the more loosely bound stellar associations and larger hierarchical structures of star formation. 
The identification of stellar associations requires a different set of methods distinct from those used to produce single-peaked compact star cluster catalogues in order to segment the light distribution over a large range of physical scales.
The refinement of such techniques is necessary to obtain a complete inventory of the youngest stellar populations and facilitate joint analysis of molecular clouds and \textsc{H\,ii} regions. 

%Motivation for a new method for identifying stellar associations.  
Previous studies of associations and star-forming regions have used a variety of methods to identify stellar associations and the larger hierarchical structures in nearby galaxies including nearest neighbours \citep[][]{gouliermis10}, the minimum spanning tree \citep[e.g,][]{cartwright04, bastian07}, two-point correlation functions \citep[][]{elmegreen18}, kernel density estimation (KDE) isopleths \citep[][]{gouliermis15,gouliermis17}, and most recently
dendrograms \citep[e.g,][]{rodriguez19,larson20}.
The nearest neighbour and two-point correlation functions provide good characterisation of smaller stellar associations while the minimum spanning tree is useful to quantify hierarchy and the degree to which structures are fractal in nature. 

We desire to trace the structures at multiple physical scales in the hierarchy in a way that easily allows for comparisons between multiple galaxies and joint analysis of molecular clouds and \textsc{H\,ii} regions. To achieve this goal, we combine techniques from both the kernel density estimation \citep{gouliermis15,gouliermis17} and dendrogram methods \citep[e.g.,][]{rosolowsky08,rodriguez19, larson20}.
Both the kernel density method and dendrograms employ isodensity contours to quantify their hierarchy.
Dendrograms identify the full branching tree of the hierarchy of any statistically significant branch in a more free-form fashion. This has its advantages of characterising the multi-scale structures and all corresponding peaks, or `leaves’, in the distribution by creating a new branch in the hierarchy every time a statistically significant structure is identified. However, the organic branching results in identified `leaves’ with a wide range of surface brightness cutoffs \citep[see][for more details]{rosolowsky08} and makes it difficult to compare structures at common surface brightness levels.
%While dendrograms are a powerful tool to trace the associations across the hierarchy, 
We have therefore chosen to develop a method that allows us to identify stellar associations and compute their hierarchical structure at distinct surface brightness cut-offs determined by physical size scales.

The kernel density method applied by \cite{gouliermis17} 
investigates the hierarchy of structures by grouping structures above surface brightness levels, while our new method groups structures by their physical scale lengths. 
%Our method instead creates hierarchy levels that are tied to different physical scale lengths, 
By relying on a fixed ladder of physical scales, our method facilitates comparison between galaxies. This allows for investigations based on physical scale lengths since disc scale heights and fragmentation lengths can differ both between and within galaxies.
The different levels of the stellar hierarchy can also be easily compared to ancillary data with a variety of resolutions while still preserving the connection to both smaller and larger structures in the hierarchy.

Our method uses a seeded watershed algorithm on local-background-filtered Gaussian kernel density maps of tracer stellar populations to systematically delineate stellar associations and hierarchical structures over a wide range of physical scales. 
Watershed algorithms have been widely applied for image segmentation in the broader field of computer vision. Pixel values in the image are treated as in a topographic map, and regions are defined by `flooding' the images starting at user defined marker positions until separate regions meet or until the boundaries reach a surface brightness equal to a specified value. The resulting regions identified by the watershed algorithm define our stellar associations at each level in the hierarchy. 

Our methods have been developed to produce complete catalogues of stellar associations using the UV-optical imaging obtained by the Physics at High Angular Resolution in Nearby GalaxieS with the \textit{Hubble Space Telescope} (PHANGS--HST) survey \citep{lee22}.  PHANGS--HST\footnote{\url{https://phangs.stsci.edu/}} is a component of the greater PHANGS\footnote{\url{http://www.phangs.org}} programme, which is building a comprehensive database to enable multi-phase, multi-scale studies of the star formation process across the nearby spiral galaxy population.  This effort is comprised of: PHANGS--ALMA, an Atacama Large Millimeter/\linebreak[0]{}submillimeter Array (ALMA) \mbox{CO(2--1)} mapping programme covering a representative sample of 74 main sequence galaxies \citep{leroy21a}; PHANGS--MUSE, a Very Large Telescope (VLT) imaging programme of 19 of these galaxies with the MUSE optical Integral Field Unit (IFU) instrument \citep{emsellem22}; and PHANGS--HST which has obtained \textit{NUV-U-B-V-I} imaging for the 38 galaxies from the PHANGS sample best-suited for studies of resolved stellar populations (including all galaxies observed with MUSE)\footnote{Galaxies are selected to be relatively face-on, avoid the Galactic plane, and have robust molecular cloud populations to facilitate joint analysis of stellar populations and molecular clouds \citep{lee22}.}. Imaging with \textit{JWST} in eight filters at $2$ to $21$~\micron{} for the 19 galaxies with the full complement of ALMA, MUSE, and \textit{HST} observations have been approved through the PHANGS--JWST Cycle~1 Treasury survey \citep[Lee et al, 2022 \textit{submitted}][PID: 2107]{}.  The ensemble of PHANGS observations yields an unprecedented database of the observed and physical properties of ${\sim}100{,}000$ star clusters, associations, \textsc{H\,ii} regions, and molecular clouds.  With these basic units of star formation, PHANGS systematically charts the evolutionary cycling between gas and stars across a diversity of galactic environments found in nearby galaxies.  

We focus the analysis on two galaxies, NGC\,3351 and NGC\,1566, to develop our methods.  We choose these galaxies because they span the distance range of most of the galaxies in the PHANGS--HST sample \citep[NGC~3351 is at 10~Mpc, while NGC~1566 is at 18~Mpc,][]{anand21} and have been previously studied by the LEGUS programme \citep{calzetti15} to allow for catalogue comparisons. 
The two galaxies exhibit different morphologies and global properties, and provide a test of our method over a large range of galactic environments and dynamical structures including bars, spiral arms, and star-forming rings. LEGUS star cluster catalogues have also been publicly released for both galaxies \citep{adamo17}, which enables comparison of the stellar associations found here to compact associations and star clusters which have been identified and studied using independent methods. LEGUS has also analysed the distribution of star formation in NGC~1566 in a hierarchical manner using KDE isopleths \citep{gouliermis17}.  

This paper is part of a series which presents the major components of the overall PHANGS--HST data products: overview of survey design, pipeline, and data products \citep{lee22}; source detection and selection of compact star cluster candidates \citep{thilker22}; aperture correction and quantitative morphologies of star clusters \citep{deger22}; star cluster candidate classification \citep{whitmore21}; deep learning for neural network classification of star clusters \citep{wei20}; stellar association identification and analysis (this paper); spectral energy distribution fitting with \textsc{cigale}:\textit{: Code Investigating GALaxy Emission \footnote{\url{https://cigale.lam.fr}}} \citep{boquien19, turner21}; and constraints on galaxy distances through analysis of the Tip of the Red Giant Branch (TRGB) as observed in the PHANGS--HST parallel pointings \citep{anand21}.

This paper is organised as follows.
We begin in Section~\ref{SEC:data} by describing the \textit{HST} data used in this analysis.  
In Section~\ref{SEC:methods}, we describe our method for identifying stellar associations. We describe our process for selecting tracer stars that are used to observationally delineate stellar associations, the creation of filtered Gaussian kernel density maps, the watershed algorithm used to define the regions bounding the associations, and the parameters involved in the definition process.
In Section~\ref{SEC:photandsed}, we describe our strategy for measuring photometry within the multi-scale associations, and for inferring ages, masses, and reddenings of the stellar associations via SED fitting using the \textsc{cigale} package. 

We present the results of this methodology in Section~\ref{SEC:results} by characterising the observed and physical properties of the ensemble population of stellar associations.  We compare our sample with the star clusters and compact associations identified by the LEGUS programme, and discuss their relationship with both H$\alpha$ observations and molecular clouds identified by the PHANGS--ALMA survey.  
Finally, we conclude by summarising the key findings of our analysis in Section~\ref{SEC:conclusions}, and discuss future work. The method presented here will be applied to the full sample of 38 PHANGS--HST galaxies to produce multi-scale association catalogues to be publicly released together with the scripts used.

%%%%%%%%%%%%%%%%%%%%%%%%%%%%%%%%%%%%%%%%%%%%%%%%%%
\section{Sample \& Data}
\label{SEC:data}

NGC~3351 is an Sb spiral galaxy with a mass of $1.88\times 10^{10}$~M$_\odot$ and a star formation rate (SFR) of $1.165$~M$_\odot$~yr$^{-1}$ \citep{leroy21a}. It has both an inner and outer star-forming ring connected by a weak bar structure.  NGC~1566 is an SABb spiral galaxy with a weak bar, two prominent star-forming spiral arms, and is ${\sim}4$~times more massive than NGC~3351. These basic properties and references are summarised in Table~\ref{TAB:gal_properties}).

\begin{figure}
    \includegraphics[width=0.23\textwidth]{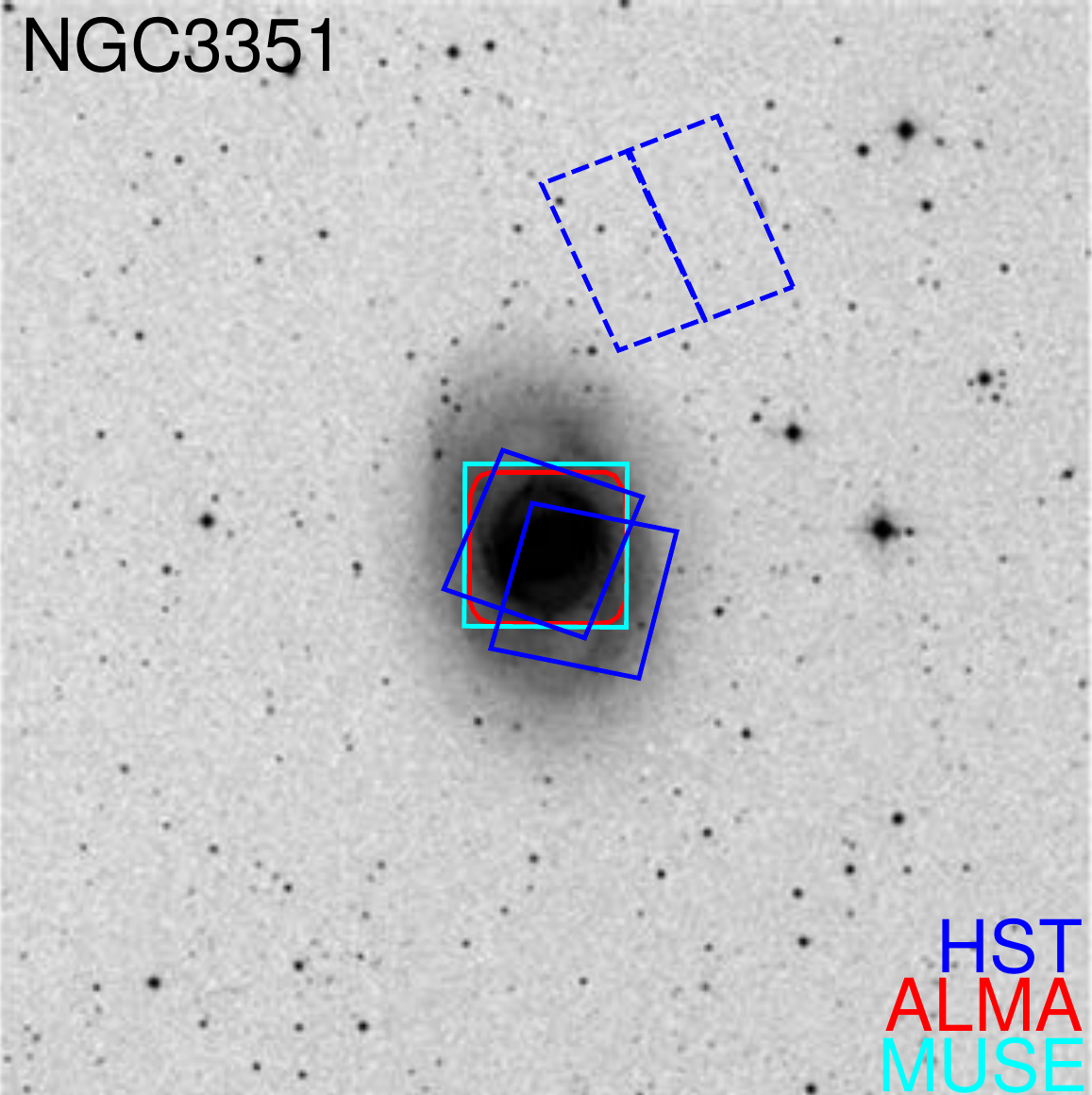}
    \includegraphics[width=0.23\textwidth]{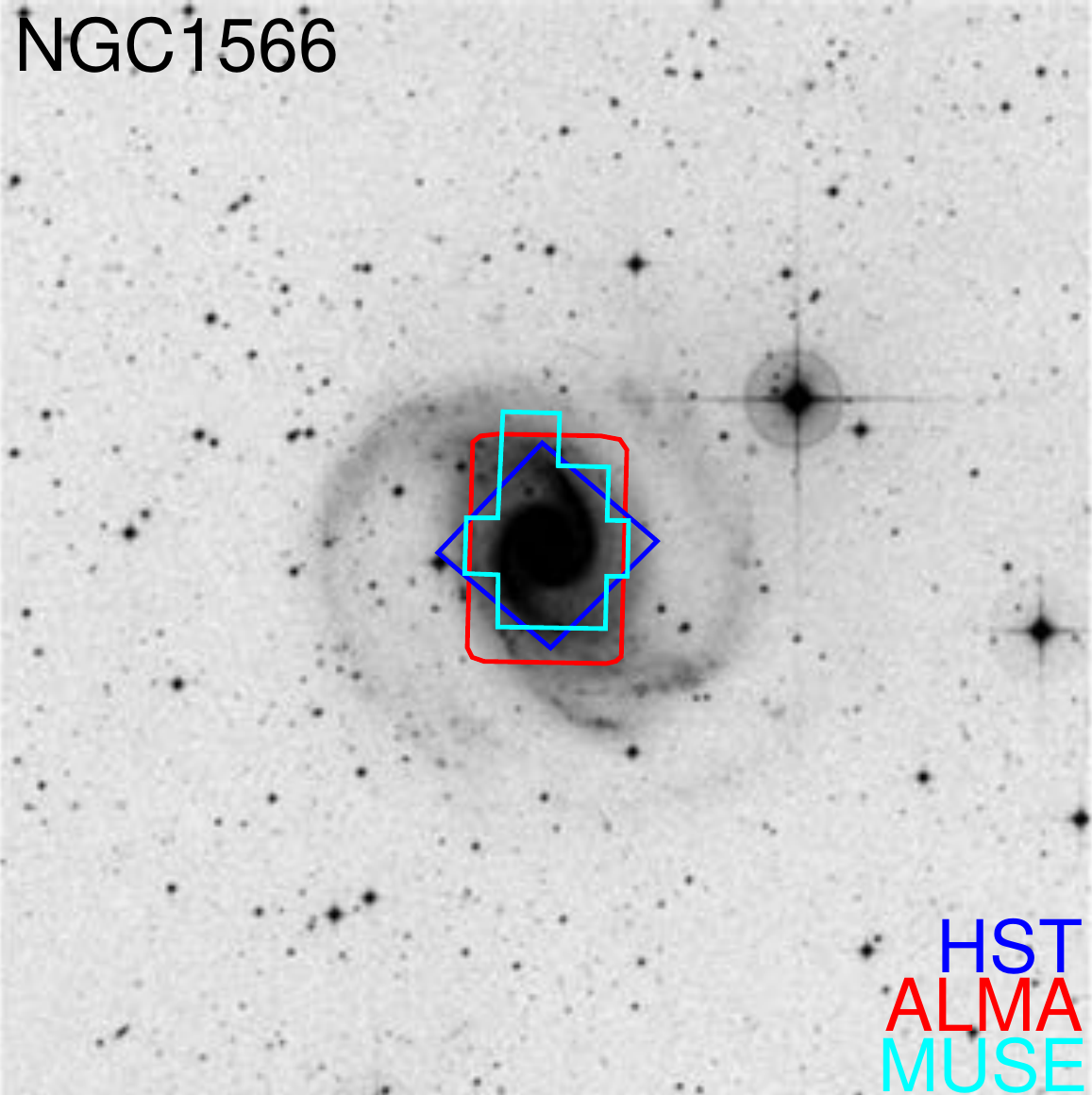}
    \caption{Footprints of the \textit{HST} (blue), ALMA (red), and MUSE (cyan) coverage for NGC~3351 and NGC~1566.  The PHANGS--HST imaging for NGC~3351 also included ACS \textit{V} and \textit{I}-band parallel observations (footprint shown as dotted blue box) to enable distance measurements \citep{anand21}. }
    \label{FIG:footprints}
\end{figure}

As described in the PHANGS--HST survey paper \citep{lee22}, all galaxies in the PHANGS--HST programme have \textit{HST} F275W~(NUV), F336W~(U), F438W~(B), F555W~(V), and F814W~(I) imaging with Wide Field Camera~3 (WFC3) or Advanced Camera for Surveys (ACS), as well as \mbox{CO(2--1)} observations with ALMA.  For NGC~3351,
two HST fields with imaging in these five bands are available for NGC~3351, one obtained by LEGUS \citep[][PID 13364]{calzetti15} in 2014, and another for the PHANGS--HST programme in 2019 \citep[][PID 15654]{lee22}.  As discussed in \cite{lee22} and \cite{turner21}, while the LEGUS observation strategy was to maximise radial coverage outward from the nucleus, PHANGS--HST aims to maximise the \textit{HST} coverage with the available PHANGS--ALMA CO mapping, which resulted in overlapping but complementary observations for NGC~3351.  NGC~1566 was observed by LEGUS in 2013.  No additional \textit{HST} observations were taken by PHANGS--HST since the available observations provided sufficient coverage of the ALMA CO map.  Figure~\ref{FIG:footprints} shows the footprints of the HST imaging together with those for the MUSE and ALMA CO maps, overlaid on Digitized Sky Survey images\footnote{The Digitized Sky Surveys were produced at the Space Telescope Science Institute under U.S.\ Government grant NAG W-2166. The images are based on photographic data obtained using the Oschin Schmidt Telescope on Palomar Mountain and the UK Schmidt Telescope. The plates were processed into the present compressed digital form with the permission of these institutions.}.  Exposure times are given in Table~\ref{TAB:exptime}.

%--------------------------------
% add total CO gas mass
% values currently from phangs_sample_table_v1p5
\begin{table*}
    \caption{Galaxy properties}
    \begin{center}
    \begin{tabular}{c|c c c c c c }
 \hline
    Name & RA & Dec & Distance & Stellar Mass & SFR & Morph\\
        &  deg & deg & Mpc & M$_{\odot}$ & M$_{\odot}$~yr$^{-1}$ &  \\
    \hline    
    NGC~3351 &	160.99065 &	11.70367 & 10& 1.88e10 (0.10 dex)& 1.2 (0.2)& Sb \\%224, 18776451560 (0.10018)
    NGC~1566	& 65.00159 & $-$54.93801 & 18 & 4.56e10	(0.10 dex) & 4.4 (0.2) &  SABb\\%204, 45615221298	(0.10095)
    \hline    
    \label{TAB:gal_properties}
    \end{tabular}
    \end{center}
    %\vspace{-10}
    \begin{tablenotes}
    \small
    \item Note: Properties of NGC~3351 and NGC~1566.
    \end{tablenotes}
\end{table*}

%--------------------------------
%--------------------------------

\begin{table*}
    \caption{\textit{HST} exposure times}
    \begin{center}
    \begin{tabular}{c|c c c c c c}
 \hline
    Name & F275W & F336W & F438W & F555W & F814W & PID \\
        & [s]   &  [s]  &  [s]  &  [s]  &  [s] &  \\
    \hline    
    NGC~3351-N & 2190 & 1110 & 1050 & 670 & 830 & 15654\\ 
    NGC~3351-S & 2361 & 1062 & 908 & 1062 & 908 & 13364\\ 
    NGC~1566 & 2382 & 1119 & 965 & 1143 & 989 & 13364\\

    \hline    
    \end{tabular}
    \label{TAB:exptime}
    \end{center}
    %\vspace{-10}
    \begin{tablenotes}
    \small
    \item Note: \textit{HST} WFC3/UVIS imaging exposure times for NGC~3351 and NGC~1566. Data for NGC~1566 are from the LEGUS programme while new data for NGC~3351 were obtained by PHANGS--HST.  
    \end{tablenotes}
\end{table*}
%--------------------------------

%The HST footprints are available at XXX and in Figure XXX in the PHANGS--HST survey paper (Lee et al. 2020, in preparation) and also in Figure XXX of Turner et al. (2020) for NGC~3351. Exposure times are given in Table XXX of the PHANGS--HST survey paper (Lee et al. 2020).  {\bf note exposure times here}

Our analysis makes use of \textsc{dolphot} source catalogues, which are produced as the first step in the overall PHANGS--HST catalogue pipeline \citep{lee22}. These \textsc{dolphot} catalogues are used as the unified base for both the PHANGS--HST cluster detection pipeline \citep{thilker22} and the hierarchical stellar structure pipeline presented in this paper. The \textsc{dolphot} photometry package \citep[v2.0;][]{dolphin02} uses point spread function (PSF)-fitting to detect and deblend sources in the high-resolution \textit{HST} images. The \textsc{dolphot} source catalogue provides PSF-fitted photometry in all available bands as well as characteristics of the source (sharpness, roundness, chi-square, signal-to-noise ratio, crowding, and quality flag) for every detection. The \textsc{dolphot} source characteristics are used to perform quality cuts on the sample as described in Section~\ref{SEC:tracer} to select the point sources used to trace the stellar associations.  \cite{thilker22} provides a detailed description of the production of \textsc{dolphot} catalogues for PHANGS--HST and lists the adopted \textsc{dolphot} parameters.

%%%%%%%%%%%%%%%%%%%%%%%%%%%%%%%%%%%%%%%%%%%%%%%%%%
\section{Multi-Scale Stellar Association Identification Method}
\label{SEC:methods}

% intro: 
\subsection{Selection of Tracer Stars}
\label{SEC:tracer}

The first step is to select point sources from the \textsc{dolphot} source catalogues to serve as `tracer' stars, the positions of which, with appropriate selection, reveal the hierarchical structure of multi-scale stellar associations.

We perform our analysis using both \textit{NUV} and \textit{V}-band light as the detection band for the \textsc{dolphot} stellar catalogues. The \textit{NUV}-band light preferentially traces the youngest stellar populations, which is critical for comparison with molecular clouds. The \textit{V}-band light facilitates comparisons to compact star cluster catalogues from both PHANGS--HST and LEGUS, which are \textit{V}-band selected \citep{adamo17,lee22,thilker22}.

For the selection of tracer stars in the \textit{NUV}-band we use the following criteria:
\begin{itemize}
  \item \textit{NUV}-band signal-to-noise $> 7$,
  \item \textit{NUV}-band sharpness$^2 < 0.15$,
  \item \textit{NUV}-band quality flag $< 4$,
  \item U-band signal-to-noise $> 3$. %(helps remove noise)
\end{itemize}
The sharpness criterion preferentially selects unresolved point sources. A perfectly fitted star has a sharpness of zero and a well fitted star in an uncrowded field has a sharpness value between $-0.3$ and $0.3$ \citep{dolphin02}. We chose a more relaxed cut-of sharpness$^2 < 0.15$ to allow for effects from crowding.  The \textsc{dolphot} quality flag criterion ensures that we only consider unsaturated sources with well determined photometry.  A corresponding minimum detection in the \textit{U}-band helps remove artefacts.

Similarly, the criteria for the tracer star selection in the \textit{V}-band are:
\begin{itemize}
  \item \textit{V}-band signal-to-noise $> 7$,
  \item \textit{V}-band sharpness$^2 < 0.15$,
  \item \textit{V}-band quality flag $< 4$,
  \item B-band signal-to-noise $> 3$, %(helps remove noise)
  \item (\textit{V}-band flux/\linebreak[0]{}\textit{V}-band sky level) $\ge 2.0$.
\end{itemize}
Since the \textit{V}-band data has more diffuse background emission, the \textit{V}-band selection has an additional criterion to remove objects with high local sky background. This ensures that \textsc{dolphot} sources included as tracer stars have significant detections above their local galactic environment. 
Additionally, foreground galactic stars are masked out and excluded from the tracer stars for both \text{NUV} and \textit{V}-band selection.

Our selection criteria result in magnitude limits of \textit{NUV} $< 25.4$ and \textit{V} $< 27.2$~mag for NGC~3351, which correspond to absolute magnitudes of $M_\mathrm{NUV} < -4.7$ and $M_\mathrm{V} < -2.8$ for a distance of 10~Mpc. Likewise, for the more distant NGC~1566 at 18~Mpc, these limits are \textit{NUV} $<24.9$ and  \textit{V} $< 27.0$~mag, and $M_\mathrm{NUV} < -6.3$ and $M_\mathrm{V} < -4.2$. The effects that the different absolute magnitude cuts have on the final results are explored in Section~\ref{SEC:NGC3351_Mag_cutoff}. We find that the different absolute magnitude limits do not affect the size or age distributions of the resulting associations but do result in fewer associations and a slight increase of the average log(M/M$_{\odot}$) by ${\sim}0.2$~dex. Since the overall change in the determined parameters is small, we chose to keep with a simple signal-to-noise selection and not apply an absolute magnitude cut on tracer stars to allow for a more complete comparison to the cluster catalogues \citep[][]{adamo17,thilker22} and data from other telescopes. 

\begin{figure}
    \includegraphics[width=0.45\textwidth]{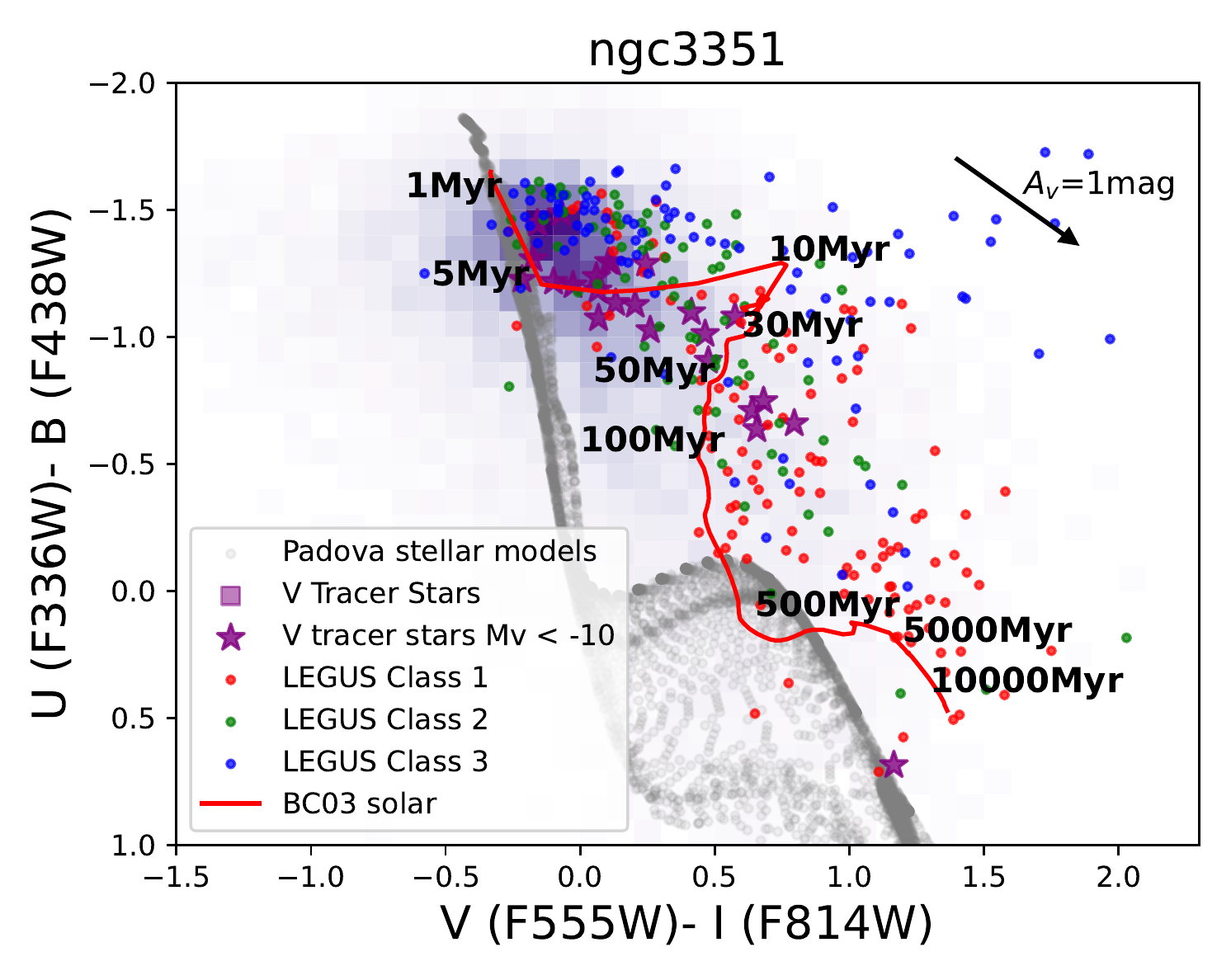}
    \caption{$UBVI$ colour--colour diagram of the \textit{V}-band tracer stars for NGC~3351. Grey points show the position of the Padova stellar models and the red line traces the path of a solar metallicity, single-aged stellar population from BC03. LEGUS objects are shown as red (Class~1 clusters), green (Class~2 clusters) and blue (Class~3 compact associations) points and tracer stars with $M_\mathrm{V}<-10$~mag are shown as purple stars. }
   \label{FIG:NUV_tracer_cc}
\end{figure}

While our criteria for the tracer stars are designed to select point-like sources, they do not guarantee that all sources are in fact individual stars. Figure~\ref{FIG:NUV_tracer_cc} shows a $UBVI$ diagram of the \textit{V}-band selected tracer stars as a purple density grid compared to both the Padova stellar model tracks\footnote{\url{http://stev.oapd.inaf.it/cgi-bin/cmd}} \citep[grey points, ][]{girardi02, marigo08} and the solar metallicity single-aged stellar population (SSP) models from  \cite{bruzual03} (BC03, red line). Most of the tracer stars lie redward of the Padova stellar tracks. Some tracers, shown as purple stars, lie beyond the Humphrys--Davidson limit of $M_\mathrm{V} < -10$~mag and are therefore too bright to be individual stars. Since the goal is to find stellar associations, it is acceptable for our purposes if some of the tracer objects are in fact double stars or even unresolved clusters. The position of LEGUS identified Class~1, 2, and~3 objects are also shown as red, green, and blue points for comparison. LEGUS Class~1 and Class~2 objects are identified as compact star clusters while Class~3 objects are multi-peaked compact associations. The LEGUS objects should be well characterised by a SSP model and lie along and to the right of the BC03 SSP track.

%---------------------------------
%
%---------------------------------

%---------------------------------
\begin{figure*}
    \includegraphics[width=0.80\textwidth]{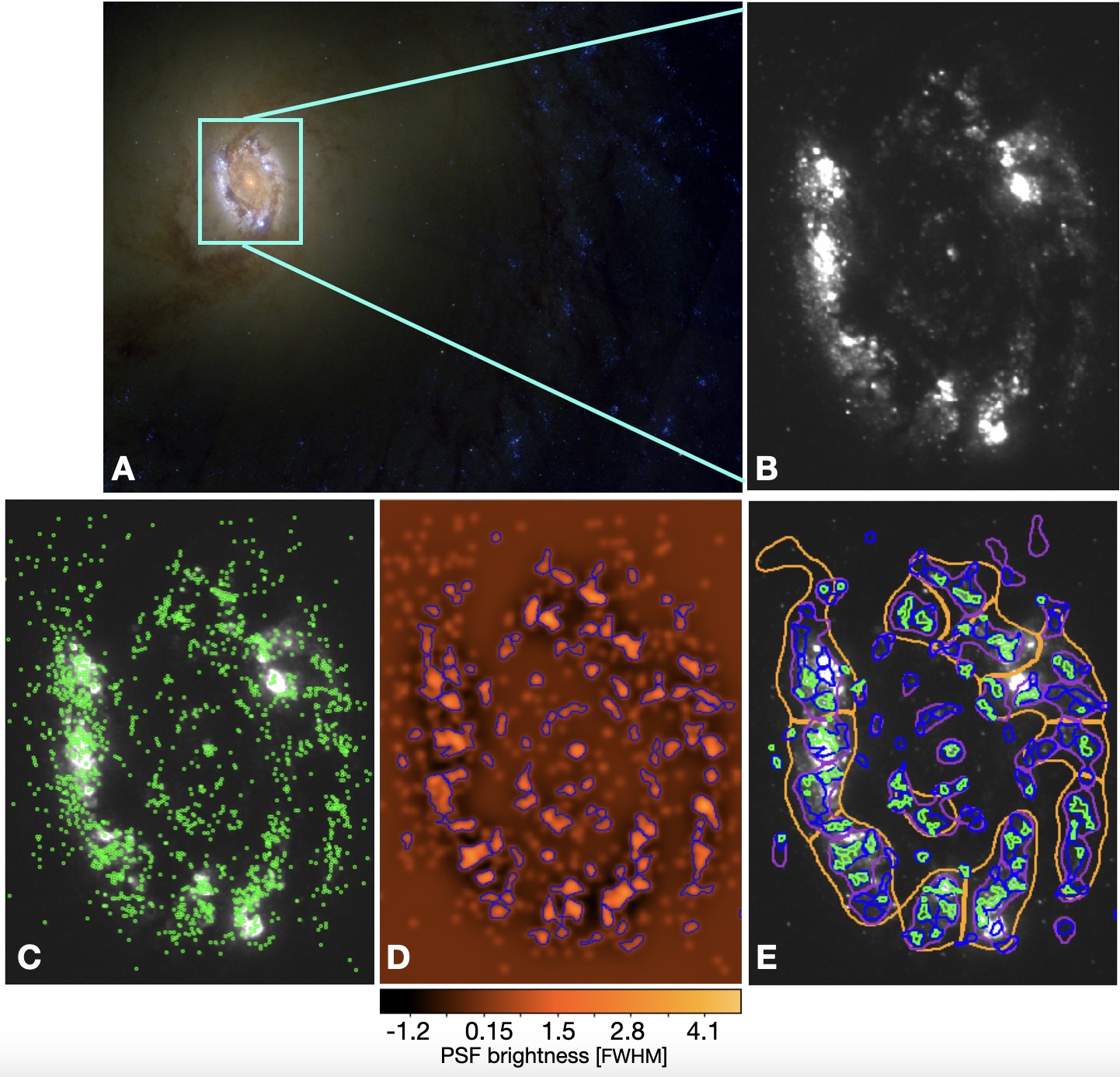}

    \caption{An illustration of our watershed procedure for identifying stellar associations over multiple scales.  \textit{(Panel~A)} An \textit{HST} three-colour image of NGC~3351 (Red: WFC3/UVIS F814W, Green: WFC3/UVIS F555W, Blue: WFC3/UVIS F438W+F336W+F275W).  \textit{(Panel~B)} The \textit{NUV} image of the center of NGC~3351. 
    \textit{(Panel~C)}The \textit{NUV} image of the center of NGC~3351 with the position of all \textit{NUV} tracer stars shown as green points 
    \textit{(Panel~D)} The \mbox{16-pc} smoothed, filtered image of the \textit{NUV} tracer stars for the center of NGC~3351. The image scale is in units of the FWHM of the PSF brightness. The cutoff threshold is set to 1 FWHM and the \mbox{16-pc} watershed regions are shown in blue.  
    % The black points designate the local maxima used for the watershed function. 
    \textit{(Panel~E)} \textit{NUV} image of the center of NGC~3351 with 8, 16, 32, and \mbox{64-pc} levels overplotted as corresponding green, blue, purple, and orange regions. At $0.0396$~arcsec~pixel$^{-1}$, one WFC3 pixel maps to $1.9$~pc~pixel$^{-1}$ at a distance of $10$~Mpc for NGC~3351.}
    \label{FIG:ngc3351tracers}
\end{figure*}
%--------------------------------
\begin{figure*}
    \includegraphics[width=0.80\textwidth]{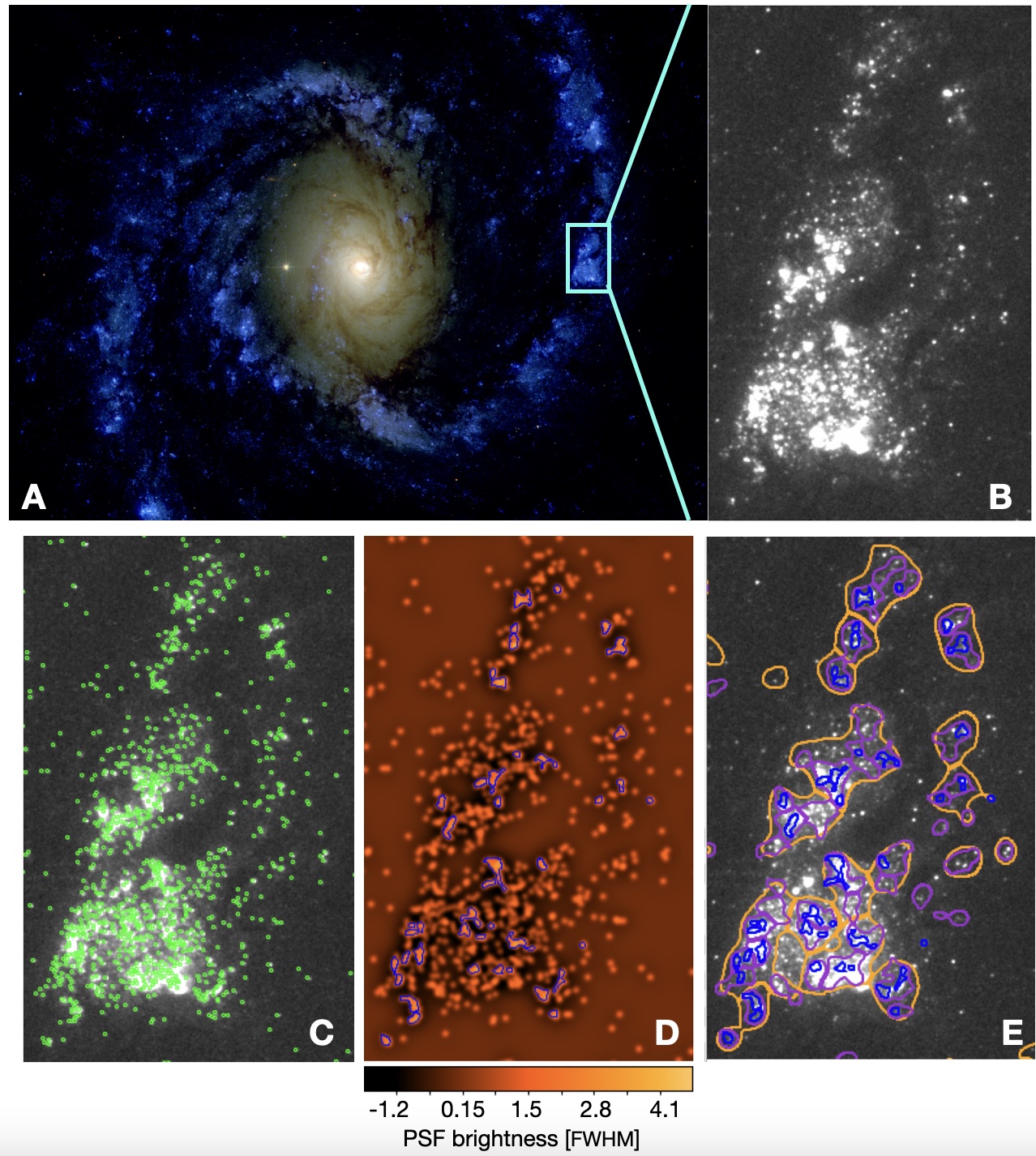}

    \caption{Same as Figure~\ref{FIG:ngc3351tracers}, except for NGC~1566. At $0.0396$~arcsec~pixel$^{-1}$, one WFC3 pixel maps to $3.5$~pc~pixel$^{-1}$ at a distance of $18$~Mpc for NGC~1566  \textit{(Panel~A)} The \textit{HST} three-colour image for NGC~1566 (Red: WFC3/UVIS F814W, Green: WFC3/UVIS F555W, Blue: WFC3/UVIS F438W+F336W+F275W).  
    \textit{(Panel~B)} For illustration, we show a zoom-in of the \textit{NUV} image on a portion of the spiral arm with a bright star-forming region in NGC~1566. 
    \textit{(Panel~C)}The \textit{NUV} image of the star-forming region in NGC~1566 with the position of all \textit{NUV} tracer stars shown as green points.   
    \textit{(Panel~D)} The \mbox{16-pc} smoothed, filtered image of the \textit{NUV} tracer stars for the zoom in region of NGC~1566. The image scale is in units of the FWHM of the PSF brightness. The cutoff threshold is set to 1 FWHM and the \mbox{16-pc} watershed regions are shown in blue.
    \textit{(Panel~E)} \textit{NUV} image of NGC~1566 with 16, 32, and \mbox{64-pc} levels overplotted as corresponding blue, purple, and orange regions.}
    \label{FIG:ngc1566tracers}
\end{figure*}

%--------------------------------

\subsection{Tracer Star Maps}
\label{SEC:TracerMaps}

Once the tracer stars are identified, they are used to create a simplified map on which we can use the watershed algorithm to identify stellar associations.
The steps to develop the tracer star maps that will undergo segmentation are as follows.

A position map of the tracer stars' positions is created with the same shape and scale as the \textit{HST} image. The position map has a value of one at the pixel position of the tracer stars and a value of zero at all other pixels. The location of \textit{NUV} tracer stars are shown for regions in NGC~3351 and NGC~1566 in figures Figures~\ref{FIG:ngc3351tracers}(C) and~\ref{FIG:ngc1566tracers}(C). Gaussian kernel density maps are then created by smoothing the position map with a Gaussian profile with different full-width-half-maximum (FWHM): 8, 16, 32, 64, 128, and 256~pc, where each successive smoothing level is two times as large as the preceding level, and where the two largest-scale levels are used to construct a high-pass filter as described below.
The different smoothing levels allow us to trace the hierarchical structure at different physical scales.
A lower limit in the smoothing level is set for each galaxy dependent on the galaxy's distance: 8~pc for NGC~3351 and 16~pc for NGC~1566. The smallest smoothing scales of 8 and 16~pc are chosen to approximately match the aperture size of 4~pixels (in radius) that is used for stellar cluster identification at the galaxies' distances given the resolution of WFC3 of $0.0396$~arcsec~pixel$^{-1}$  \citep[]{adamo17,thilker22, lee22}.  The smallest scale levels will therefore enable identification and comparison of stellar associations that overlap in size with objects in compact cluster catalogues.   
These smoothed tracer star maps are then processed using a high-pass filter to emphasise the detailed structure at the chosen smoothing scale. The high-pass filter is applied by subtracting a tracer star map that is smoothed with a larger Gaussian kernel.
For example, the filtered \mbox{16-pc} tracer star map, as shown in Figures~\ref{FIG:ngc3351tracers}(D) and~\ref{FIG:ngc1566tracers}(D), is created by subtracting a map which is smoothed to 64~pc (a scale 4~times larger). The high-pass filter acts as a local background subtraction, filtering out the underlying distribution of stars in the image and emphasising the structure at each smoothing scale. 
These smoothed, high-pass filtered tracer star maps are then segmented using the watershed algorithm, as shown in Figures~\ref{FIG:ngc3351tracers}~(E) and~\ref{FIG:ngc1566tracers}~(E) and described in the next Section \ref{SEC:WatershedParams}, to define stellar associations at different scales.
In Appendix~\ref{SEC:appendix}, we compare the use of two different high-pass filters and select the 4~times high-pass filter for our sample as this filter step produces a tighter distribution of region sizes at all smoothing scales. 

%---
\subsection{Watershed Parameters}
\label{SEC:WatershedParams}

We use the seeded watershed routine from the publicly available python package scikit-image (\textsc{skimage.segmentation.watershed}) to define our stellar associations. We choose this particular watershed routine since it allows the user to control both the identification of the marker positions, which serve as the starting points from which regions are `flooded' or grown, as well as to provide an image mask for pixels to consider when creating the segmented regions. 

We develop a procedure to determine the marker positions and to produce the image mask using two free parameters:
\begin{itemize}
    \item Peak threshold: the level above which marker positions are identified.
    \item Edge threshold: the level above which image mask pixels are valid (have value of~1).
\end{itemize}

We tie the values of both parameters to the characteristics of a single object on the smoothed, filtered tracer star maps.  We define the peak threshold to be $1.5$~times the maximum value for a single object to ensure that there will be at least two point-like sources per region.  We use an edge threshold equal to the surface `brightness' at the FWHM of a single object.  Both of these values vary as a function of: (i)~the distance of the galaxy, since the FWHM of the Gaussian kernels used to smooth the images are defined in parsecs, and (ii)~the smoothing level itself. 

We define our marker positions as the local maxima of the smoothed, filtered tracer star map using the routine \textsc{skimage.feature.peak\_local\_max}. The local maxima function requires both a minimum distance and threshold value to define the local maxima. We set the minimum allowed distance between maxima equal to the FWHM of a single object.

\section{Photometry \& SED Fitting Methods} 
\label{SEC:photandsed}
\subsection{Stellar Association Photometry}
In this section we describe how we perform the photometry in the stellar associations which we use for the determination of physical properties. Since the stellar associations are often irregular in shape and by definition contain multiple stellar PSFs, we can not assume that the standard aperture correction calculated for the compact cluster catalogues is appropriate for these regions \citep{deger22}. We therefore preform both regional pixel photometry and PSF photometry to test which procedure best recovers the total flux of the associations at all scale levels.

We first measure the flux of the stellar associations using standard regional pixel photometry. Regional photometry is preformed by measuring the flux inside the stellar association regions and calculating the local background from an annulus around the region. Flux from all pixels within the regions are included in the photometry, therefore flux from compact clusters and diffuse emission within the aperture will also add to the regional photometry. We then measure PSF photometry of the stellar associations using \textsc{dolphot} identified sources.

The PSF photometry of the stellar associations is found by summing \textsc{dolphot} PSF photometry from all the \textsc{dolphot} sources contained within the region. The photometric error for the associations is calculated by propagating the \textsc{dolphot} photometry errors for the individual sources. While the definition of the associations only uses \textsc{dolphot} sources that pass the tracer star selection (see Section~\ref{SEC:tracer}), the PSF photometry uses all \textsc{dolphot} sources. Bright foreground stars are masked out from the tracer stars and are excluded from the association regions. Therefore, the total number of \textsc{dolphot} sources contained within an association and used for photometry can be greater than the number of selected tracer stars used to define the associations.  We discuss this later in Section~\ref{sec:nsources}.  No additional selection criteria are applied to the \textsc{dolphot} sources. \textsc{dolphot} only fits sources above the background with PSF-fitted photometry, therefore no additional background subtraction is required for the PSF photometry.

\textsc{dolphot} is iteratively fitted 5~times. The highest signal-to-noise sources are fitted in the first pass, while the subsequent passes allow for fainter sources to be fitted in the residuals. Not all \textsc{dolphot} sources are point sources, but the inclusion of all \textsc{dolphot} sources in the photometry allows for the total flux of the region to be recovered. Since the same initial \textsc{dolphot} catalogue is used for both the stellar association pipeline and the single-peaked compact cluster pipeline, it is possible for compact clusters to be included as part of a larger stellar association.  As we will show in Section~\ref{SEC:ages_masses}, the observed and physical properties appear to be reasonable (e.g., the colours of the associations track that of a single-aged stellar population) and sufficient for this proof of concept demonstration.

\begin{figure*}
    \includegraphics[width=0.9\textwidth]{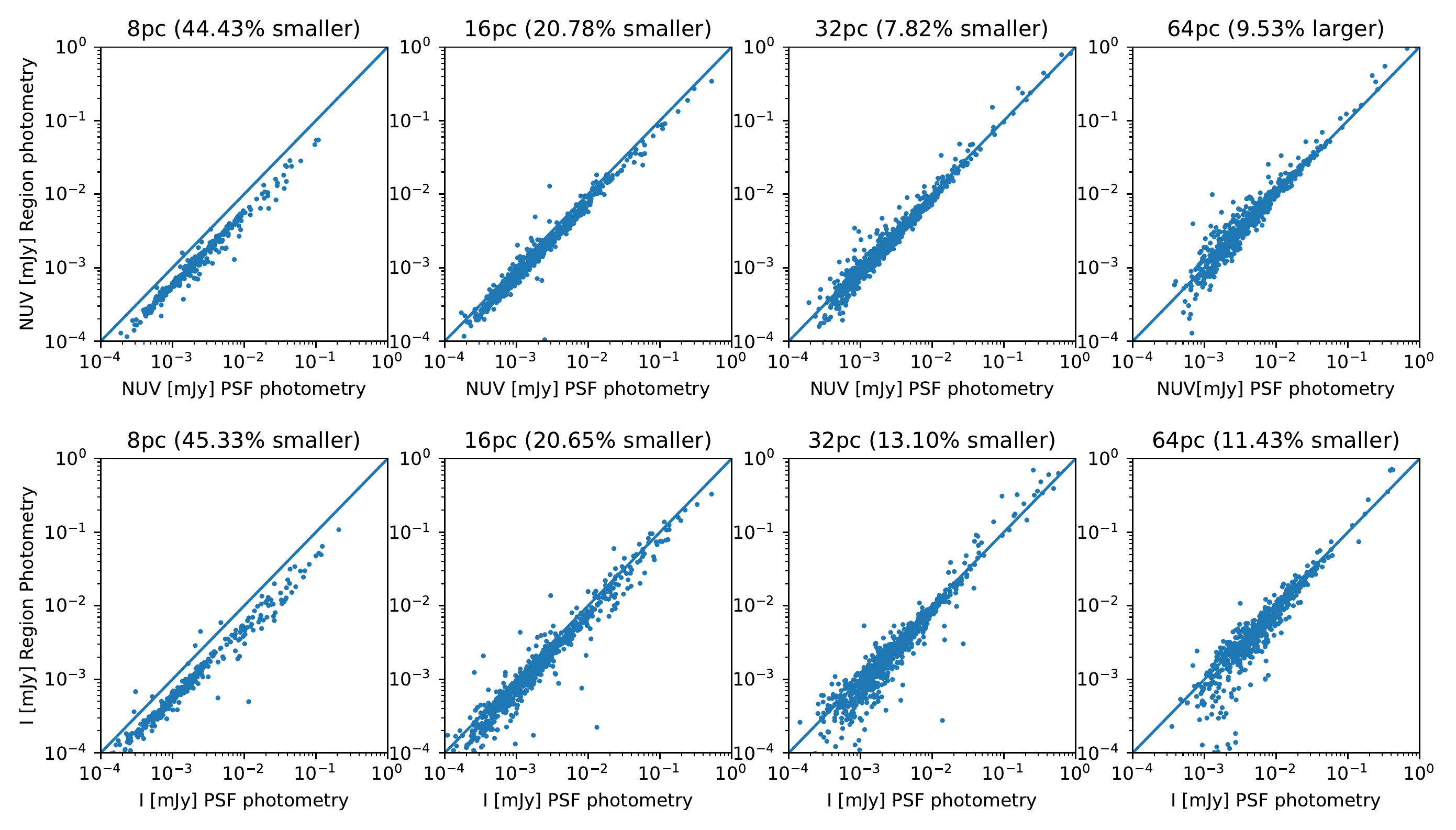}
    \caption{Comparison of regional and PSF photometry of stellar association for NGC~3351 for all scales (8 to 64~pc). Top row shows the comparison of \textit{NUV}-band photometry and bottom row shows the comparison for \textit{V}-band photometry. The title of each sub-figure states the average difference between the regional and PSF photometry.}
    \label{FIG:phot_compare}
\end{figure*}

We compare the the regional sky subtracted pixel photometry to the PSF photometry of the stellar associations. Figure \ref{FIG:phot_compare} shows how both the \textit{NUV} and \textit{V}-band regional and PSF photometry compare for NGC~3351. On average, the regional photometry is less than the PSF photometry. The difference is largest at the \mbox{8-pc} scale level with the regional photometry being on average $45$-per-cent smaller than the PSF photometry. The difference decreases as the regions get larger at larger scale levels. By the \mbox{64-pc} level, the regional photometry is $\sim$ $11$-per-cent different from the PSF photometry. 

Since the \textsc{dolphot} photometry is PSF fitted and represents the total flux for that object, it does not require an aperture correction or local background subtraction which can be difficult in crowded environments. There is no aperture correction for the irregular shapes of the stellar associations used in the regional photometry. The regional photometry of the smaller regions at 8 and 16~pc are therefore most impacted by loss of flux due to small apertures. We will therefore use the PSF photometry in our analysis for the rest of the paper.

%---------------------------------
%--------------------------------

\subsection{SED Fitting}
To estimate ages, masses, and reddenings of the stellar associations, the \textsc{cigale} SED fitting package \citep{boquien19} is used with five-band \textit{NUV, U, V, B, I} photometry of the regions as described in the previous section.  Augmentations to \textsc{cigale} have been made to enable fitting of single-aged stellar populations as described in \cite{turner21}. Here, all of the results presented are based on the best-fitting SED corresponding to the minimum reduced chi-squared value, and we do not introduce additional weighting with priors on the age or mass distributions.  The fitting is based on the single-aged population synthesis models of \cite{bruzual03}, assuming solar metallicity and a \cite{chabrier03} initial mass function (IMF; with standard mass limits of $0.1{-}100$~M$_{\odot}$), and no addition of nebular emission.  The \cite{cardelli89} extinction curve with $R_\mathrm{V} = 3.1$ is used and a maximum $E(B{-}V) = 1.5$ is imposed. A more detailed description of the assumptions in the modelling can be found in \cite{turner21}.

\section{RESULTS: Stellar association properties}
\label{SEC:results}

Now that we have developed a process for characterising stellar associations, let us examine the resulting ensemble properties of these structures in NGC~3351 and NGC~1566.  We begin by describing the properties of the associations based upon the \textit{NUV} tracer stars, at all four scale levels studied (8, 16, 32, and 64~pc).  We discuss both observed and physical properties of these structures. We then discuss differences in the samples of selected structures when the \textit{V}-band tracer stars are used instead.  The statistics for the sizes, number of regions, and number of point sources contained in the regions at each scale level are tabulated in Tables~\ref{TAB:ngc3351_properties} and~\ref{TAB:ngc1566_properties}. Similar tables for the estimated ages, masses, and extinctions for stellar associations at each scale level are given in Tables~\ref{TAB:ncg3351_age_mass} and~\ref{TAB:ncg1566_age_mass}.

\subsection{Number of regions identified and their sizes}

The different scale levels used in our watershed procedure 
trace the hierarchical structure of star formation from the more compact associations at 8 and 16~pc, to the larger scale stellar associations at 32 and 64~pc. 
We find that our process identifies on the order of a few hundred up to over three thousand associations depending on the galaxy and scale level (Tables~\ref{TAB:ngc3351_properties} and~\ref{TAB:ngc1566_properties}).  For NGC~1566 we find around two times more regions than NGC~3351 at each scale level.  This potentially follows from the higher SFR values quoted by \cite{leroy21a} (see Table~\ref{TAB:gal_properties}) for NGC~1566 . For each galaxy, as the smallest scale level traces the densest peaks of the stellar hierarchy, it also has the fewest regions, while the number of regions is highest at the \mbox{32-pc} level. 
As can be seen in Figures~\ref{FIG:ngc3351tracers}~and~\ref{FIG:ngc1566tracers}, while smaller \mbox{8-pc} and \mbox{16-pc} regions generally lie within the larger \mbox{32-pc} structures, not every \mbox{32-pc} region will have substructure at smaller scales. This is due to the requirement that at least two tracer stars be within a region so some of the \mbox{32-pc} regions are too diffuse to have substructure at the smaller scales.
The \textit{V}-band selected tracer stars also result in more detected associations at each scale level with ${\sim}1.5{-}2$ times the number of associations detected with \textit{V}-band tracer stars than are detected with \textit{NUV}-band tracer stars. We will return to examine the possible causes of these differences in Section~\ref{SEC:nuv_v}.

\begin{figure}
    \includegraphics[width=0.44\textwidth]{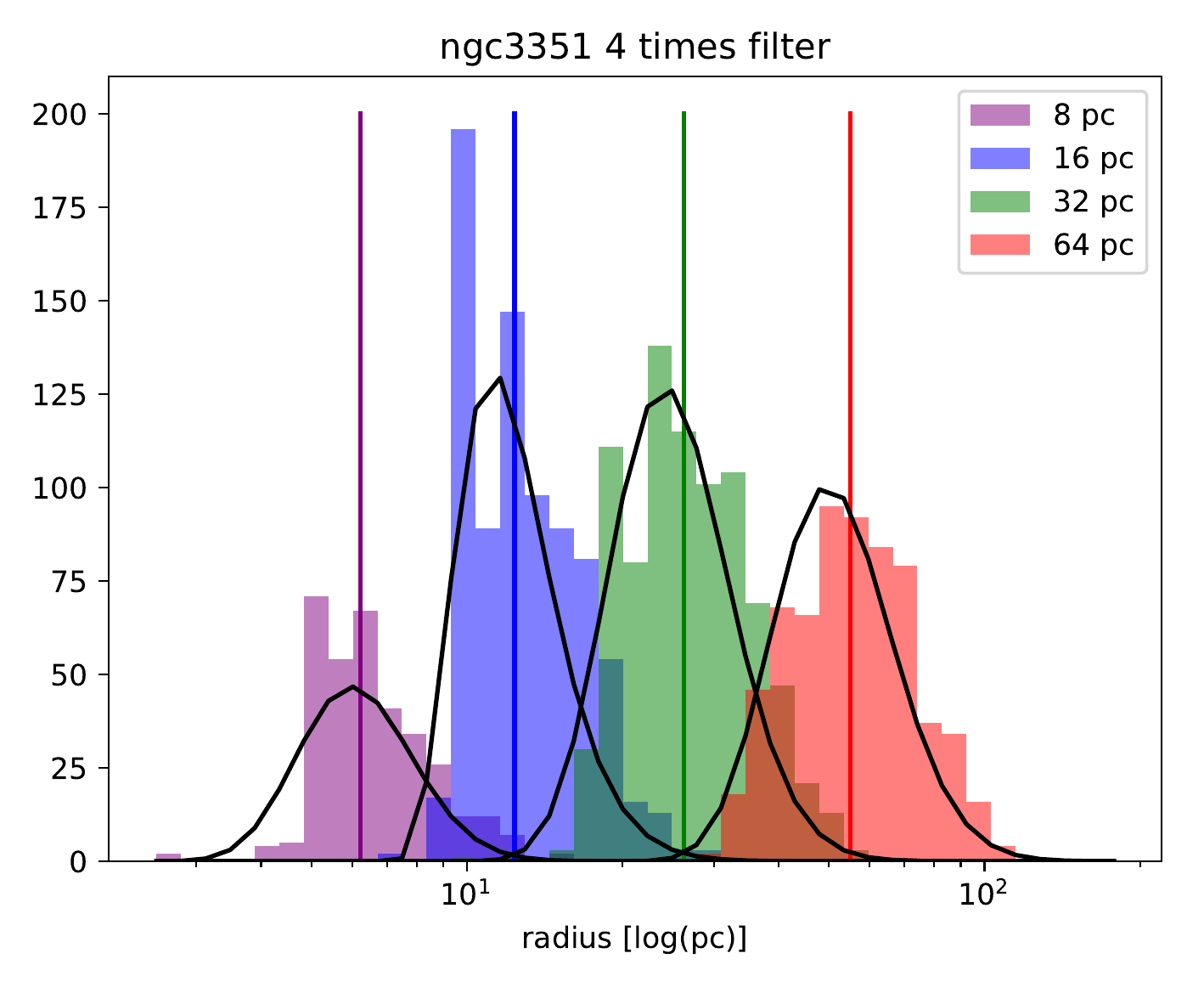}
    \includegraphics[width=0.44\textwidth]{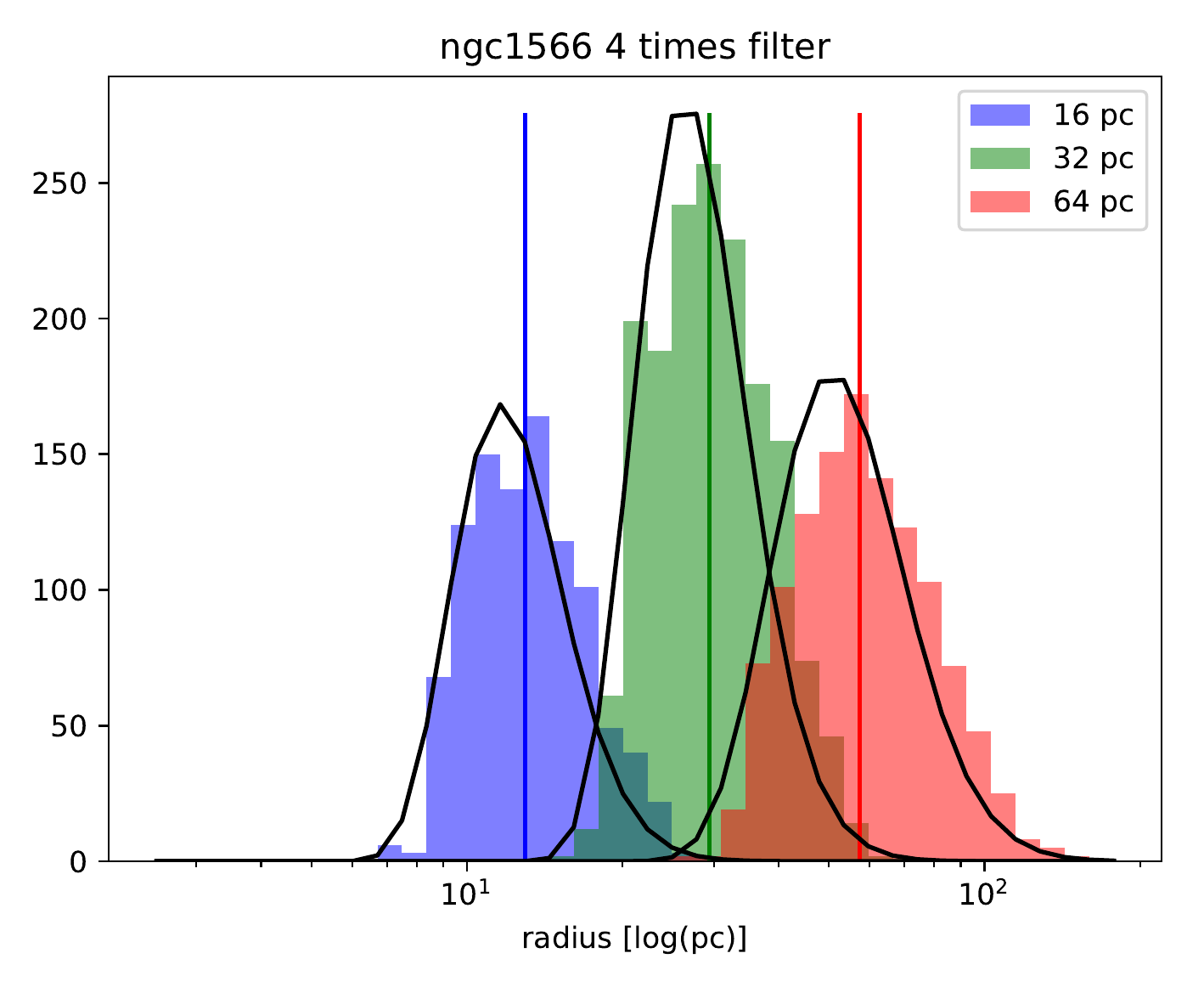}
    \caption{Distribution of effective radii for \textit{NUV} identified stellar associations in NGC~3351 and NGC~1566. A log-normal fitted to each distribution is overplotted in black and the median radius is marked with a vertical line. The median values are also reported in Tables~\ref{TAB:ngc3351_properties} and~\ref{TAB:ngc1566_properties}.}
    \label{FIG:size}
\end{figure}

Each scale level is optimised to identify associations of different sizes. Here we use the region's effective radius as a measure of the size, where the effective radius is the radius of a circle with an area equal to that of the association. 
The histograms of the effective radii for each scale are shown in Figure~\ref{FIG:size}. The sizes peak near the defining Gaussian FWHM used to create the tracer star maps for each scale level (8, 16, 32, and 64~pc) and are approximately log-normal as shown by the fitted log-normal distributions in black. The high-pass filtering described in Section~\ref{SEC:TracerMaps} results in distributions that are fairly distinct with overlap only in the wings of the distribution. For example, the \textit{NUV}-band \mbox{16-pc} scale level sample of stellar associations for NGC~3351 has a median size of $12.4$~pc with quartiles of $10$ and $15$~pc. The distribution also exhibits a tail that extends up to $35$~pc.  The next scale level at $32$~pc has a median size of $26$~pc with quartiles of $22$ to $32$~pc. While this overlaps the small tail from the previous scale level, the peaks of the distributions are distinct and shows that our method is performing as intended.

%--------------------------------
%region properties
\begin{table*}
    \caption{NGC~3351 region properties}

    \begin{center}
    \begin{tabular}{ c | c c c c c c c}
    \multicolumn{8}{c}{NGC~3351 \textit{NUV}-band tracer stars} \\
    \hline
    scale level &  &  & nRegions & radius & nTracer & n\textsc{dolphot} & $M_\mathrm{V}$ [mag]\\
    {[pc]} & [pix] & [arcsec] &  & med (.25Q,.75Q) [pc] & med  (.25Q,.75Q) & med (.25Q,.75Q) & med (.25Q,.75Q)\\
    \hline

    8 & 4.16 & 0.165 & 334 & 6.23 (5.53,7.66) & 3 (3,6) & 4 (3,6) & -6.8 (-7.88,-6.01)  \\
    16 & 8.33 & 0.33 & 807 & 12.4 (10.3,15.3) & 4 (3,7) & 7 (4,11) & -6.86 (-7.8,-6.1) \\
    32 & 16.7 & 0.66 & 838 & 26.1 (21.6,32.4) & 5 (3,8) & 16 (10,26) & -7.06 (-7.97,-6.35) \\
    64 & 33.3 & 1.32 & 641 & 55.2 (45.2,67.4) & 7 (4,13) & 49 (29,74) & -7.71 (-8.47,-7.03) \\

    \hline
    \multicolumn{8}{c}{NGC~3351 \textit{V}-band tracer stars} \\
     \hline

    8 & 4.16 & 0.165 & 557 & 6.03 (5.31,7.11) & 3 (2,5) & 4 (3,5) & -6.16 (-7.22,-5.33) \\
    16 & 8.33 & 0.33 & 1324 & 11.9 (10.2,14.7) & 3 (2,6) & 6 (4,10) & -6.27 (-7.28,-5.52)\\
    32 & 16.7 & 0.66 & 1379 & 25.7 (21.1,31.3) & 5 (3,8) & 14 (9,24) & -6.63 (-7.51,-5.95)\\
    64 & 33.3 & 1.32 & 962 & 53.9 (43.8,67.1) & 8 (4,13) & 42.0 (27.0,70.8) & -7.36 (-8.24,-6.65) \\ 
    \hline
    \label{TAB:ngc3351_properties}
    \end{tabular}
    \end{center}
    \vspace{-15pt}
    \begin{tablenotes}
    \small
    \item Note: Region properties for NGC~3351 associations using both NUV and \textit{V}-band tracer stars. [columns 1--3] FWHM of the scale level. [column~4] number of identified associations. [column~5] median, first, and third quartiles for the association radii. [column~6] median number of included tracer stars. [column~7] median number of included \textsc{dolphot} sources. [column~8] median absolute \textit{V}-band magnitude of the associations determined from \textsc{dolphot} photometry 
    \end{tablenotes}
    
\end{table*}

%2625 total regions
%Number of NUV Tracers Stars selected: 11812
%Number of \textit{V}-band Tracers Stars selected: 20472
%--------------------------------

%--------------------------------
%region properties
\begin{table*}
    \caption{NGC~1566 region properties}
    \begin{center}
\begin{tabular}{ c | c c c c c c c}
\hline
\multicolumn{8}{c}{NGC~1566 \textit{NUV}-band tracer stars} \\
 \hline
scale level &  &  & nRegions & radius & nTracer & n\textsc{dolphot} & $M_\mathrm{V}$ [mag] \\
{[pc]} & [pix] & [arcsec] &  & med (.25Q,.75Q) [pc] & med (.25Q,.75Q) & med (.25Q,.75Q) & med (.25Q,.75Q) \\
\hline
16.0 & 4.63 & 0.183 & 983 & 12.9 (10.9,15.5) & 5.0 (3.0,7.0) & 5.0 (3.5,8.0) & -8.08 (-8.85,-7.33)   \\ 
32.0 & 9.26 & 0.367 & 1655 & 29.3 (24.4,35.7) & 6.0 (4.0,11.0) & 14.0 (9.0,22.0) & -8.45 (-9.32,-7.7)    \\
64.0 & 18.5 & 0.733 & 1175 & 57.4 (46.5,72.0) & 8.0 (4.0,16.0) & 34.0 (20.0,60.0) & -8.92 (-9.84,-8.13)  \\
\hline
\multicolumn{8}{c}{NGC~1566 \textit{V}-band tracer stars} \\
 \hline
16.0 & 4.63 & 0.183 & 2155 & 12.2 (10.1,15.0) & 4.0 (3.0,7.0) & 5.0 (3.0,8.0) & -7.59 (-8.36,-6.86)  \\ 
32.0 & 9.26 & 0.367 & 3150 & 27.3 (23.0,33.7) & 6.0 (3.0,10.0) & 11.0 (7.0,19.0) & -8.01 (-8.81,-7.24) \\
64.0 & 18.5 & 0.733 & 1995 & 55.4 (45.2,69.3) & 8.0 (4.0,17.0) & 29.0 (16.0,51.0) & -8.57 (-9.48,-7.66) \\  
\hline
\label{TAB:ngc1566_properties}
%2971 total regions
%Number of NUV Tracers Stars selected: 11812
%Number of \textit{V}-band Tracers Stars selected: 20472
\end{tabular}
\end{center}
    \vspace{-15pt}
    \begin{tablenotes}
    \small
    \item Note: Region properties for NGC~1566 associations using both NUV and \textit{V}-band tracer stars. [columns 1--3] FWHM of the scale level. [column~4] number of identified associations. [column~5] median, first, and third quartiles for the association radii. [column~6] median number of included tracer stars. [column~7] median number of included \textsc{dolphot} sources. [column~8] median absolute \textit{V}-band magnitude of the associations determined from \textsc{dolphot} photometry 
    \end{tablenotes}

\end{table*}
%--------------------------------

%---------------------------------
%--------------------------------

\subsection{Number of point sources}
\label{sec:nsources}

We can count the number of point sources enclosed in each association. We define the number of stars (point sources) in two ways: the number of stars that meet the selection criteria and are used to define the association (tracer stars), and the total number of \textsc{dolphot} sources enclosed in each association. It is a requirement that each region contains at least two tracer stars in the filtered image, see Section~\ref{SEC:tracer} for details. For both galaxies, the median number of tracer stars per region ranges from 3 to 8 as the scale level increases from 8 to 64~pc.

For NGC~3351, the median number of \textsc{dolphot} sources contained in the stellar associations vary from 4~sources at the smallest scale level of 8~pc, up to 49~sources at the \mbox{64-pc} scale level. For NGC~1566, the median number of contained \textsc{dolphot} sources ranges from 5 to~34.

\subsection{Colors}
Here, we examine the photometric colours of our sample of stellar associations using the UBVI colour--colour diagram. In past works, the \textit{UBVI} colour--colour diagram has been used extensively for comparison of star cluster populations to single-age stellar population (SSP) model tracks \citep[e.g.,][]{chandar10b, adamo17, cook19, deger22}.
Determining whether the photometric properties of the structures are consistent with single-age populations in this way provides a sanity-check on the validity of our watershed method for selecting stellar associations, and on whether to proceed with SED fitting with SSP models. We also compare to composite stellar populations (CSPs) that are formed from exponentially declining star formation rate, $\mathrm{SFR}(t) \sim \mathrm{exp}(-t/\tau)$, with both $\tau$ of $2$ and $4$~Myr. 

\begin{figure}
    \includegraphics[width=0.48\textwidth]{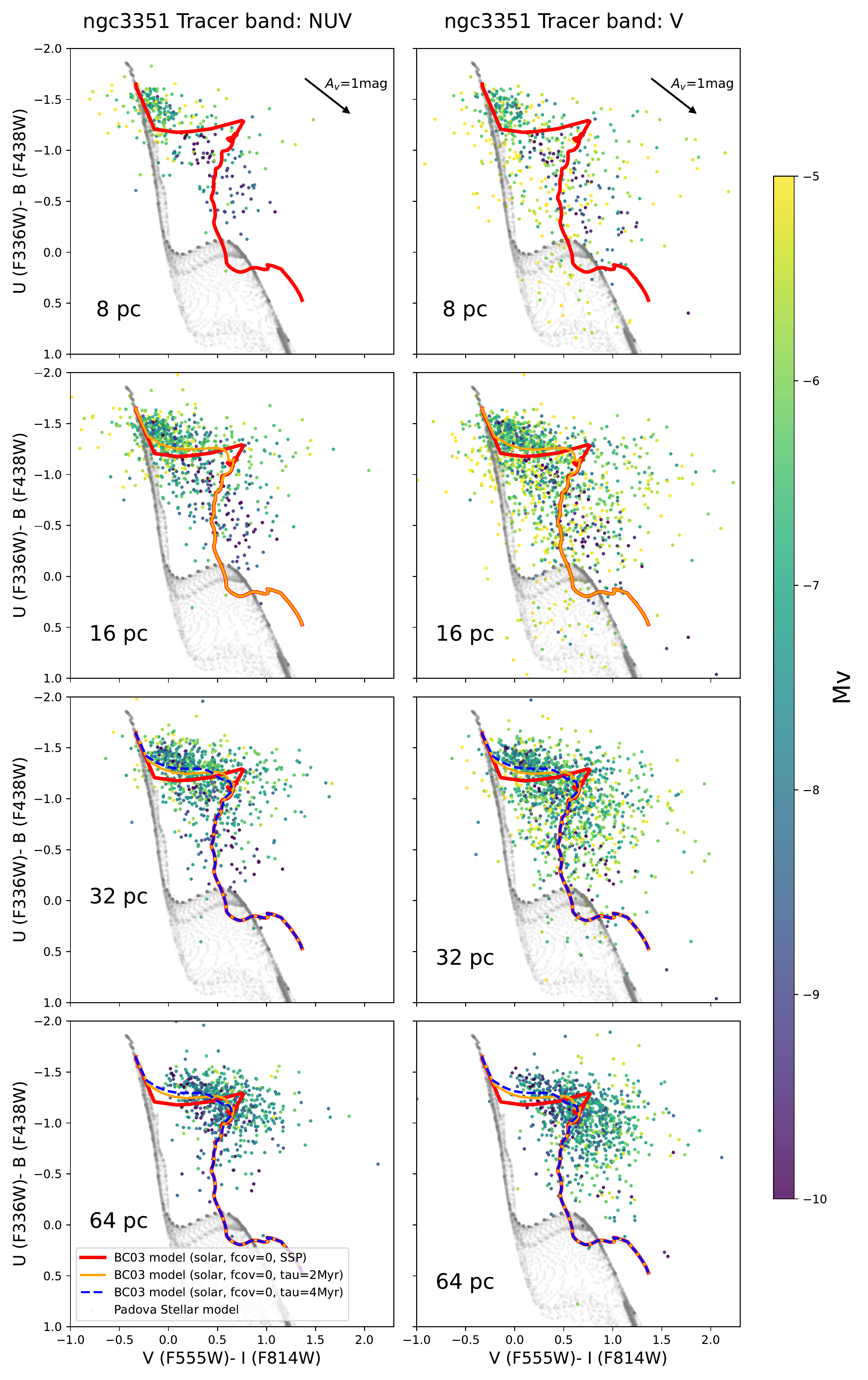}
    \caption{Left: NGC~3351 UBVI colour--colour diagrams for stellar associations selected in the \textit{NUV}-band (left) and \textit{V}-band (right)for four scale levels (8, 16, 32, and 64~pc). The associations are coloured according to their absolute \textit{V}-band magnitude ($M_\mathrm{V}$). The Padova stellar models  are shown as grey points and the solar metallicity, $f_\mathrm{cov} = 0$ evolutionary tracks of synthetic stellar populations from BC03 are overplotted on the associations. The red line traces the path of a single-aged stellar population, while the yellow and blue lines trace composite stellar populations with $\tau = 2$ and $\tau = 4$, respectively.
    }
    \label{FIG:NGC3351_NUV_V_cc}
\end{figure}

\begin{figure}
    \includegraphics[width=0.48\textwidth]{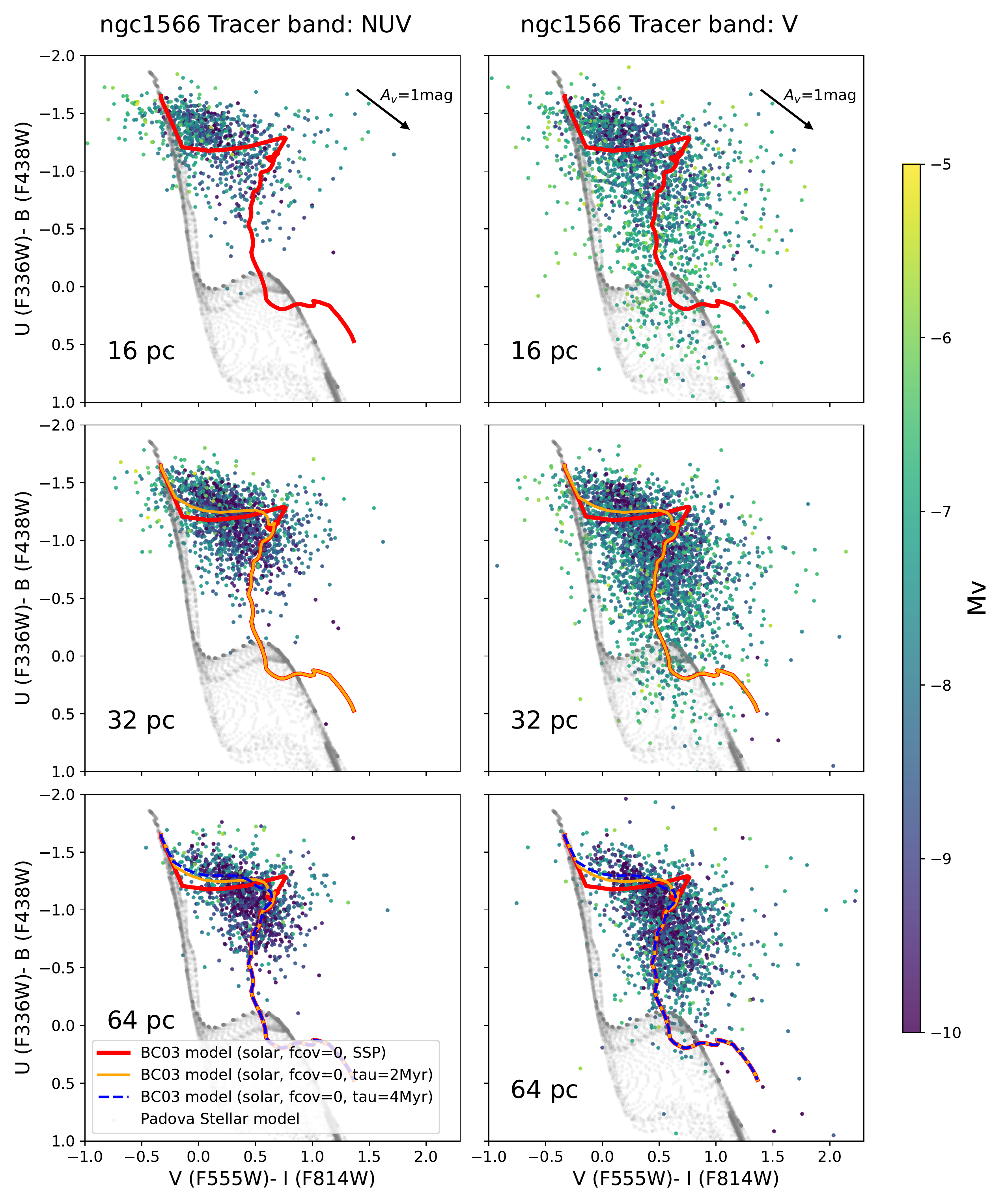}
    \caption{Same as Figure \ref{FIG:NGC3351_NUV_V_cc}, but for NGC~1566
    }
    \label{FIG:NGC1566_NUV_V_cc}
    
\end{figure}

In Figures~\ref{FIG:NGC3351_NUV_V_cc} and~\ref{FIG:NGC1566_NUV_V_cc}, we show the UBVI colour--colour diagrams for associations in NGC~3351 and NGC~1566.  For NGC~3351, we plot the stellar associations selected at each of the four scale levels, while for NGC~1566, stellar associations selected at the \mbox{8-pc} scale level are omitted due to the greater distance of the galaxy.  The figures include two columns for each galaxy, one for the NUV-selected associations and a second for the \textit{V}-band associations.  In each panel, we overplot an SSP model track with solar metallicty and an escape fraction of zero ($f_\mathrm{cov} = 0$) from BC03 in red. The CSP tracks with $\tau$ of $2$ and $4$~Myr in yellow and blue are overplotted on the larger scale levels from 16 to 64~pc .  

Overall, the colours do tend to fall along the SSP model track.  At the smaller scale levels (8 and 16~pc), there is scatter to the left of the SSP model track into the region of the colour--colour diagram occupied by individual stars. This scatter is due to the low luminosities of the regions and indicates that they would be better modeled by a stellar track than a full stellar population.
Scatter to the right of the SSP model track has been shown to be the result of stochastic sampling of the initial mass function for low mass stellar populations \citep[e.g.,][]{fouesneau12, hannon19, whitmore20}.  A single bright red star in a stellar association can pull the colour of that region redward away from the SSP track. 

The locus of stellar associations near the beginning of the stellar tracks lies slightly to the right of the SSP track for both galaxies at all scale levels and is consistent with what has been seen previously in cluster studies \citep{chandar10b}. This mismatch from the models is thought to be due to the inability of the models to reproduce the correct colour at the youngest ($<3$~Myr) ages. 
Contributions from composite stellar population to the associations could partly explain the shift of the associations at young ages. However, it is difficult to disentangle potential dust effects that could also account for the reddening colours \cite{deger22}. 

We find that for both galaxies, the stellar associations selected at larger scale levels (32 to 64~pc) have less scatter, as can be expected since larger regions tend to contain more mass overall and stochastic sampling has less of an impact on the resulting observed properties.  The loci are tightest at the \mbox{64-pc} scale and shift redward compared to the smaller scale levels. 
 
It is possible that an SSP is not the best estimate for the larger scale level associations. It would be difficult to randomly coordinate a single burst of star formation over a 32 or \mbox{64-pc} region and we could expect the larger scale levels to deviate from a pure SSP track. However, the SSP and CSP tracks only deviate slightly in the first 10~Myr of star formation and a simple variation of an exponential star formation history is not enough to explain the observed difference from the 16 to \mbox{64-pc} scale levels.  The redward shift at the \mbox{64-pc} scale level could be the result of older stellar populations being captured at a higher relative rate as the aperture size increases. A quantitative analysis of the increase in median age of associations with increasing scale level is provided in the next section based upon ages computed from SSP SED fitting. Since the change in the stellar tracks from SSP to CSP is small and fails to explain the shifts in the colour distribution from the 16 to \mbox{64-pc} scale levels, for the rest of the paper we focus our analysis on the BC03 SSP track. Further study of more complex star formation histories will need to be done in the future.

We also compare the resulting associations at each scale level selected by the different tracer bands. The \textit{NUV}-band selected stellar associations consistently have a tighter, bluer distributions of colours. The \textit{V}-band selected associations have a similar locus at blue colours but also extend to redder colours. This extension to redder colours generally follows along the BC03 SSP track and is apparent at all scale levels for both galaxies.

%---------------------------------
%--------------------------------

%---------------------------------
%--------------------------------

%--------------------------------
\subsection{Ages and Masses}
\label{SEC:ages_masses}
\begin{figure}
    \includegraphics[width=0.235\textwidth]{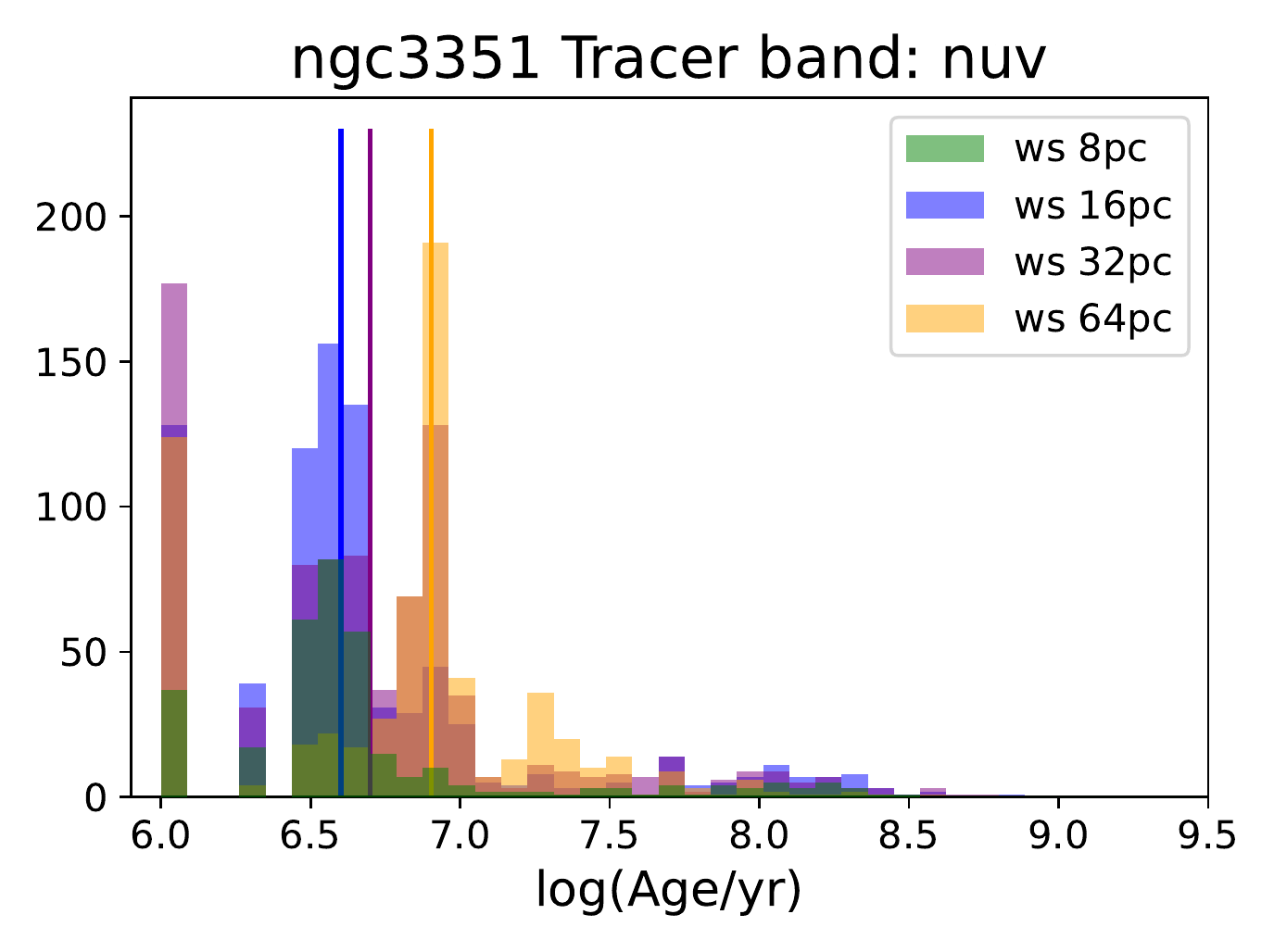}
    \includegraphics[width=0.235\textwidth]{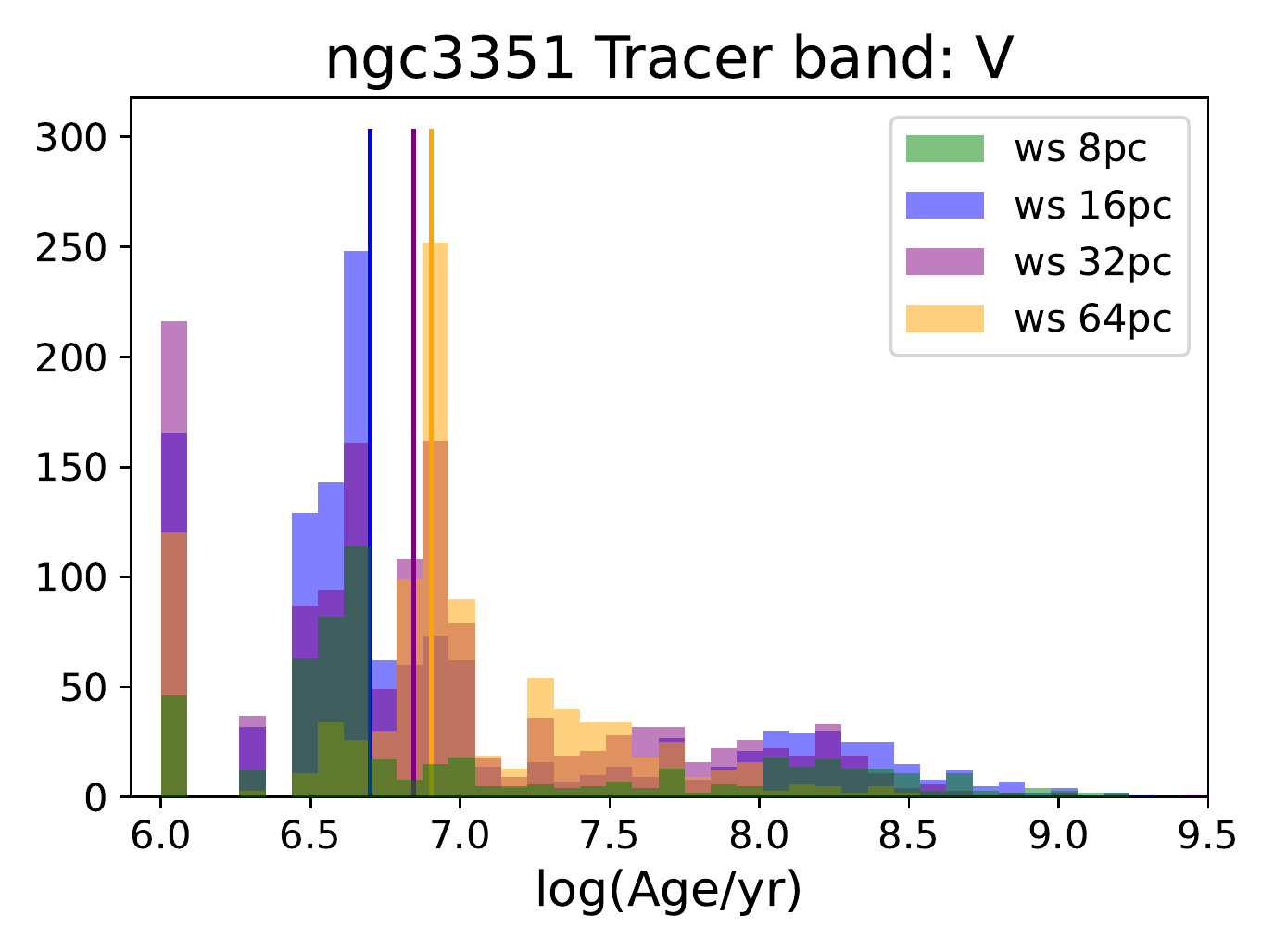}

    \includegraphics[width=0.235\textwidth]{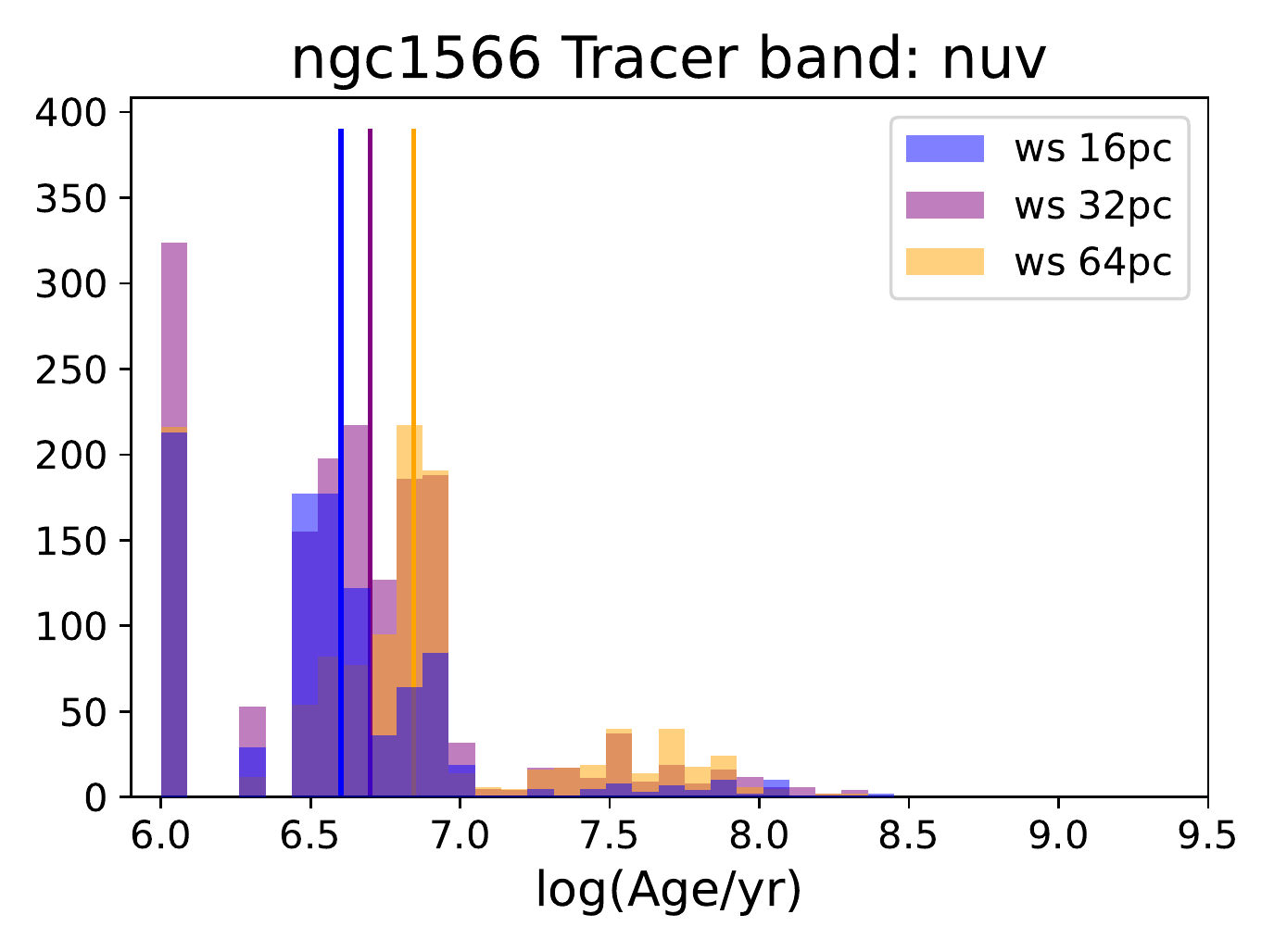}
    \includegraphics[width=0.235\textwidth]{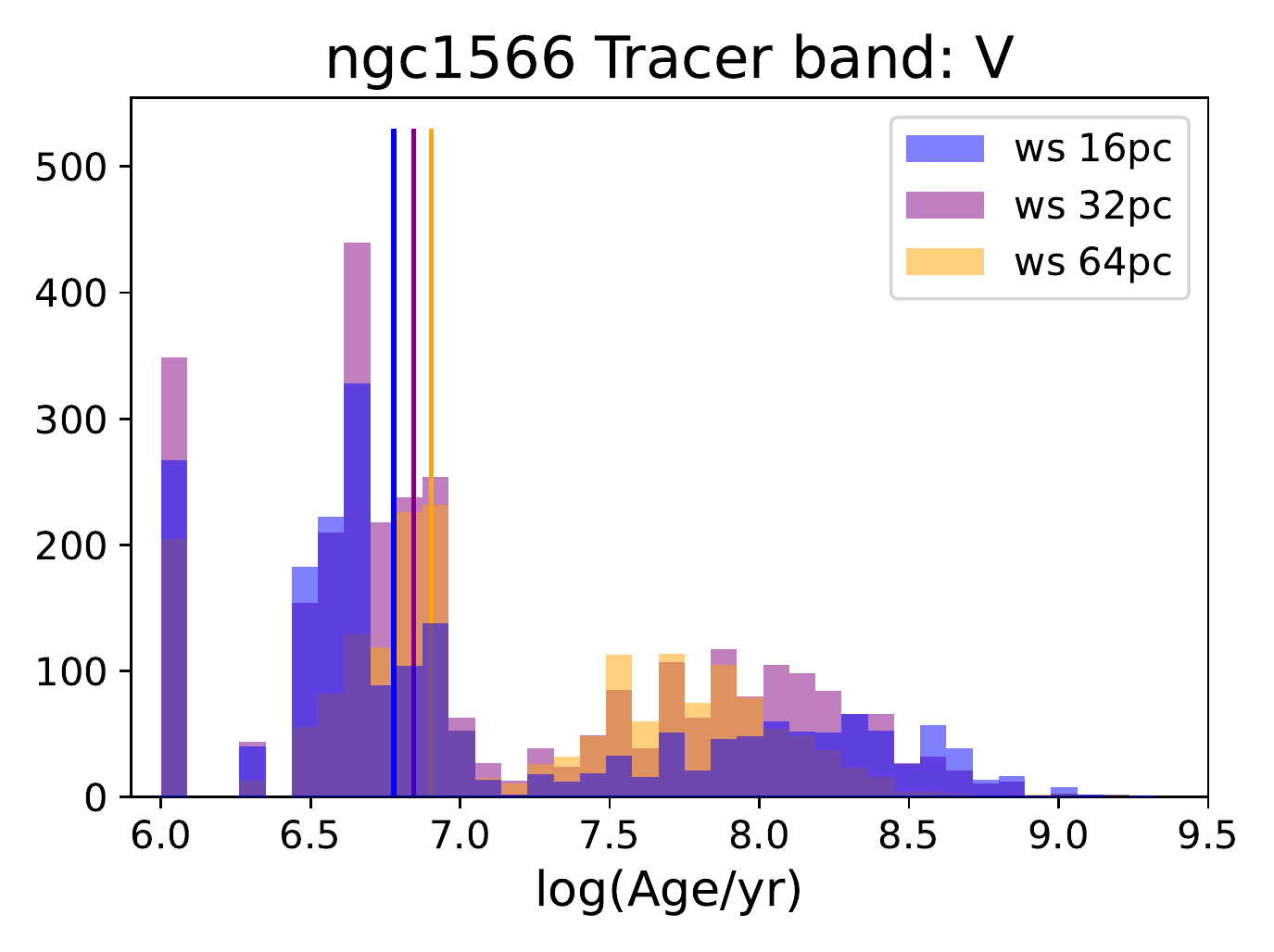}
    
    \caption{Estimated ages for NGC~3351 (top) and NGC~1566 (bottom) using both the NUV and \textit{V}-band traced watershed associations (8~pc in green, 16~pc in blue, 32~pc in purple, and 64~pc in orange).}
    \label{FIG:Age_hist_all}
\end{figure}

%--------------------------------
We compute the ages and masses of our stellar associations using the SED fitting code \textsc{cigale} with five-band \textit{NUV, U, B, V, I} photometry as input (Section~\ref{SEC:photandsed}).  For both galaxies, we find that the distributions of the ages shifts to older values at larger scale levels as can be seen in Figure~\ref{FIG:Age_hist_all}. This is also reflected in the shift in the loci to redder values at larger scale levels in the colour--colour diagrams, see Figures~\ref{FIG:NGC3351_NUV_V_cc} and~\ref{FIG:NGC1566_NUV_V_cc} as discussed above.  As the stellar associations increase in size with larger scale levels, the median ages increase from ${\sim}6.6$ to $6.9$ log(age/yr) from the \mbox{16-pc} to the \mbox{64-pc} scale level for both NGC~3351 and NGC~1566. 

The distribution of ages in Figure~\ref{FIG:Age_hist_all} appears to be somewhat bimodal, especially for NGC~1566, with peaks around log(age/yr) = 6.7 and 8.0. Part of this effect is due to a well known \citep[e.g.][]{chandar10b} artifact that produces a gap between log(age/yr) = 7.0 and 7.5. This effect is caused by the shape of the BC03 synthetic stellar population tracks, which have sharp bends and loops in this age range (see Figure~\ref{FIG:NGC3351_NUV_V_cc}), effectively shadowing certain ages and shuffling the estimates ages to younger or older ages. Another effect that may artificially overpopulate the diagram with older objects is the inclusion of some fraction of class 1 clusters, since the dolphot program occasionally breaks single, slightly extended objects into two or more separate sources, which are subsequently included as potential associations. This primarily affects the \mbox{16-pc} scale associations, since the \mbox{64-pc} associations generally have much larger numbers of tracer stars, hence diluting the inclusion of just a few objects from older clusters. This probably explains the trend toward younger ages (i.e., log(age/yr) $\sim$7.5--8.0 instead to 8.0--9.0) for the scale \mbox{64-pc} objects. 

The \textit{V}-band selection results in samples of stellar associations that have ${\sim}0.15$~dex older median age for $8$ to $32$~pc scale levels.  While the median of the age distribution for the \textit{V}-band selection is similar to the \textit{NUV}-band selection, the \textit{V}-band selection finds a larger tail of older/aged associations for both galaxies with log(age/yr) ranging from $7.25$ to~$9.0$ (see Figure~\ref{FIG:Age_hist_all}).  The \mbox{64-pc} scale level averages over the extremes of the smaller scales and has a tighter distribution of ages that is similar for both \textit{NUV} and \textit{V}-band selections.

\begin{figure}
    \includegraphics[width=0.47\textwidth]{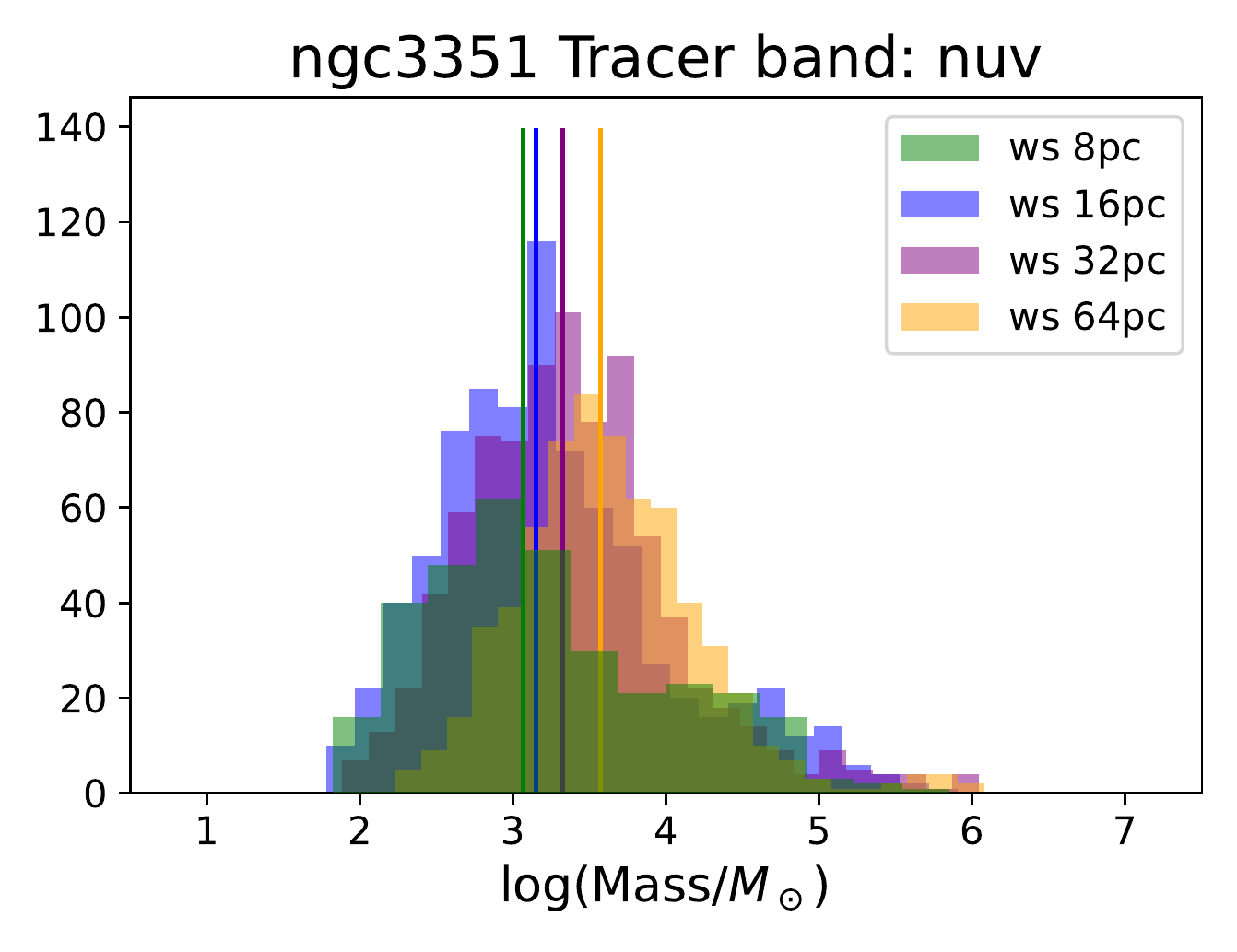}
    \includegraphics[width=0.47\textwidth]{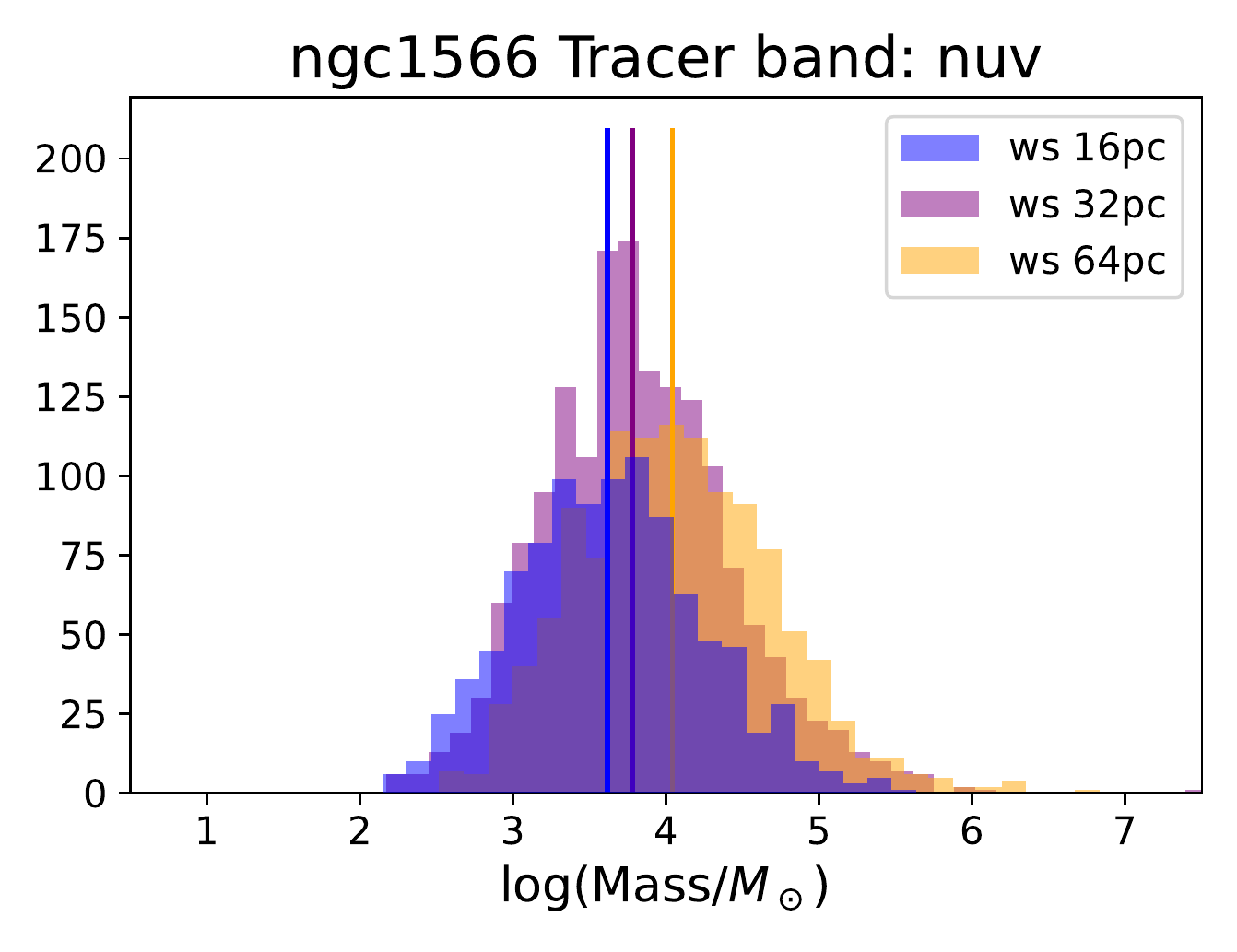}
    
    \caption{Estimated masses for NGC~3351 (top) and NGC~1566 (bottom) using \textit{NUV}-band traced watershed associations (8~pc in green, 16~pc in blue, 32~pc in purple, and 64~pc in orange).}
    \label{FIG:Mass_hist_all}
\end{figure}

For larger associations, a larger number of \textsc{dolphot} sources are included in each region as previously discussed.  Naturally, as the scale level increases, the median mass of the associations also increases (see Figure~\ref{FIG:Mass_hist_all}).  The masses of our structures range from $\log(\mathrm{M}_{\odot}) = 3{-}7$, and the median mass for associations increases by $1.6$~times from the \mbox{16-pc} to the \mbox{64-pc} scale level for both galaxies. 

%1.9 times more massive
At each scale level, the masses of the \textit{NUV}-band selected stellar associations in NGC~1566 are on average $2.9$~times higher and the and the \textit{V}-band selected are on average $3.8$~times higher than those in NGC~3351. The median masses for NGC~3351 are relatively unchanged regardless of whether the associations are \textit{NUV} or \textit{V}-band selected. The NGC~1566 \textit{V}-band selected associations are about 1.3 times more massive than those selected with \textit{NUV}.However, NGC~1566 is about twice as distant as NGC~3351. The tracer star determination is distance dependent since there is a signal to noise cutoff resulting in different absolute magnitude limits resulting in the absolute magnitude of the tracer stars in NGC~1566 being $\sim$1.5 times larger than those of NGC~3351. We therefore investigate how a different absolute magnitude limits affect the results in the next section. \ref{SEC:NGC3351_Mag_cutoff}.

%---------------------------------
%--------------------------------
%---------------------------------
%--------------------------------

\begin{figure}
    \includegraphics[width=0.48\textwidth]{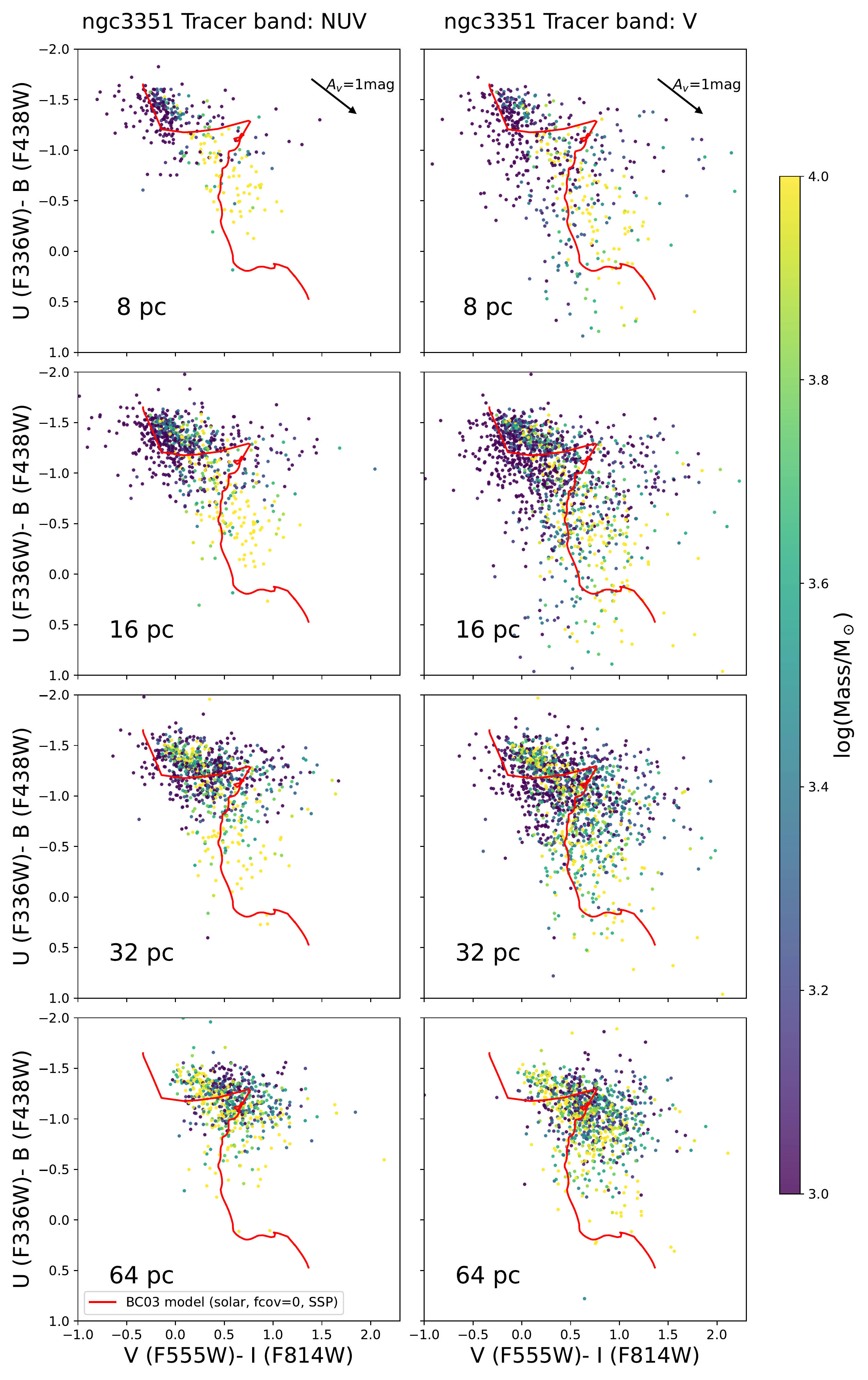}

    \caption{The NGC~3351 regions are colour-coded by their mass. The masses of the watershed regions are estimated using chi-squared fitting to the SED. The left column shows the results of watershed regions created using the \textit{NUV}-band (F275W) for the tracer stars. The right column contains results of watershed regions created using the \textit{V}-band (F555W) for the tracer stars.}
    \label{FIG:NGC3351_cc_cigMass}
\end{figure}

\begin{figure}
    \includegraphics[width=0.48\textwidth]{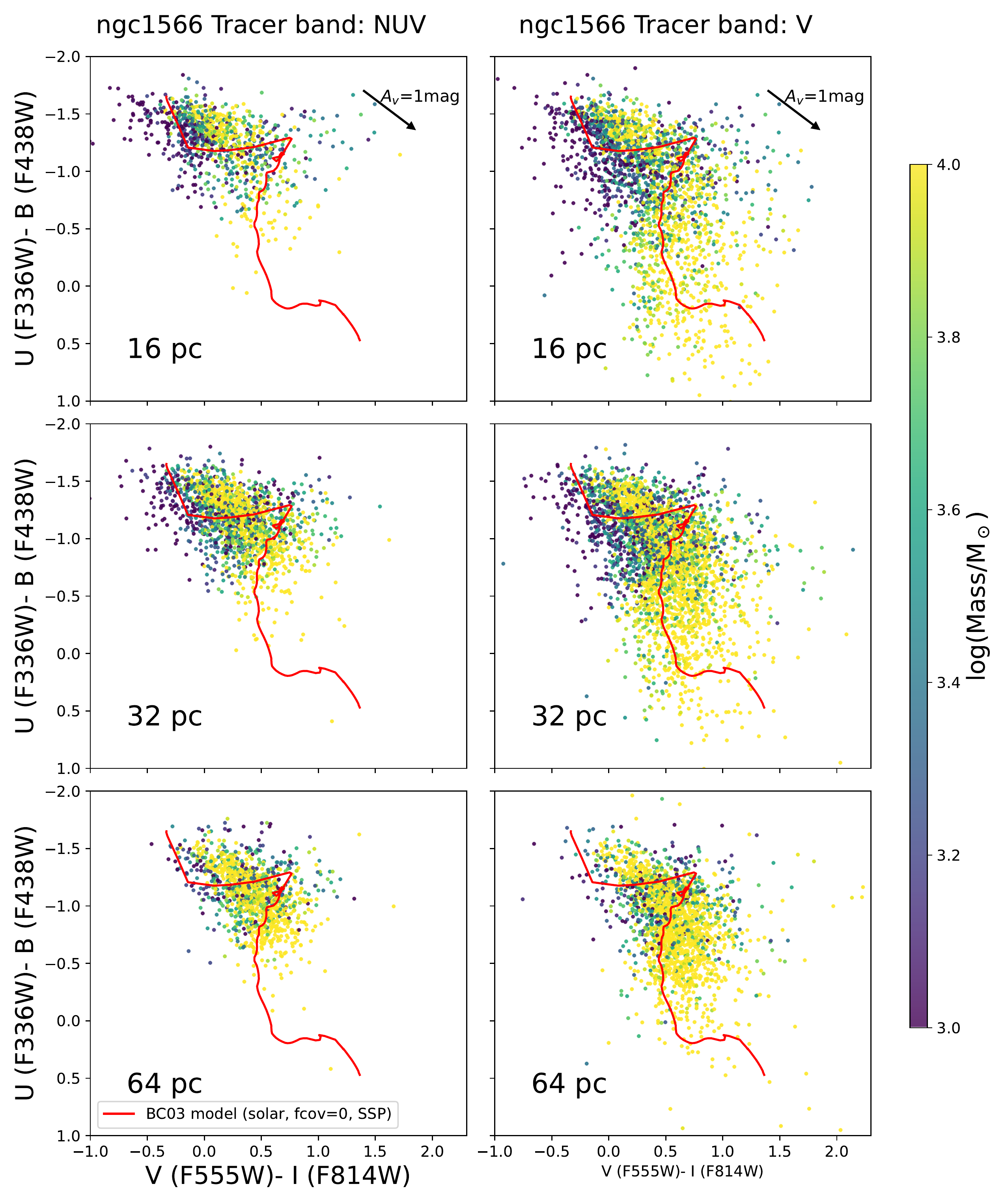}
    \caption{The NGC~1566 regions are colour-coded by their mass. The masses of the watershed regions are estimated using chi squared fitting to the SED. The left column shows the results of watershed regions created using the \textit{NUV}-band (F275W) for the tracer stars. The right column contains results of watershed regions created using the \textit{V}-band (F555W) for the tracer stars.}
    \label{FIG:NGC1566_cc_cigMass}
\end{figure}

Figures~\ref{FIG:NGC3351_cc_cigMass} and~\ref{FIG:NGC1566_cc_cigMass} show colour--colour diagrams for both NGC~3351 and NGC~1566 which are similar to the ones shown previously but now coloured by the regions' masses. Here, we see that the regions that deviate the most from the SSP track are below $\log (M_\star) \sim 4.5$. This is consistent with the regions which have the lowest mass also having the least number of \textsc{dolphot} sources and therefore are more susceptible to stochastic sampling. The most massive regions $\log (M_\star) > 4.5$ tend to follow the SSP track.

%---------------------------------
%--------------------------------
%---------------------------------
%--------------------------------

%--------------------------------
\begin{table}
    \caption{NGC~3351 \textsc{cigale} derived properties}
    \begin{center}
    \begin{tabular}{c|c c c}
    \hline
    \multicolumn{4}{c}{\textit{NUV}-band \textsc{cigale} properties} \\
 \hline
    scale level & Age & Mass & $E(B{-}V)$\\
    parsec  & log$_{10}$(yr) &log$_{10}$(M$_{\odot}$) & \\
      &med  (.25Q,.75Q)& med  (.25Q,.75Q) & med  (.25Q,.75Q)  \\
    \hline    
8~pc	& 6.6 (6.48,6.7) & 3.07 (2.62,3.72) & 0.155 (0.08,0.25)\\
16~pc	& 6.6 (6.48,6.78) & 3.15 (2.72,3.64) & 0.18 (0.08,0.28)\\
32~pc	& 6.7 (6.48,6.9) & 3.33 (2.9,3.74) & 0.18 (0.06,0.3)\\
64~pc	& 6.9 (6.6,7.0) & 3.57 (3.24,3.97) & 0.14 (0.06,0.26)\\
    \hline 
    \hline
    \multicolumn{4}{c}{V-band \textsc{cigale} properties}\\
    \hline
    8~pc & 6.7 (6.6,7.69) & 3.13 (2.7,3.6) & 0.18 (0.07,0.31) \\
    16~pc & 6.7 (6.6,7.26) & 3.17 (2.73,3.59) & 0.2 (0.08,0.32) \\
    32~pc & 6.85 (6.6,7.28) & 3.33 (2.92,3.71) & 0.21 (0.09,0.33)\\
    64~pc & 6.9 (6.85,7.28) & 3.54 (3.22,3.91) & 0.17 (0.08,0.28)\\
    \hline
    \end{tabular}
    \label{TAB:ncg3351_age_mass}
    \end{center}
    \vspace{-10pt}
    \begin{tablenotes}
    \small
    \item Note: \textsc{cigale} derived properties for both the \textit{NUV} and \textit{V}-band selected associations in NGC~3351: [col2] median ages. [col3] median masses. [col4] median attenuation.
    \end{tablenotes}
\end{table}

%--------------------------------
%--------------------------------
\begin{table}
    \caption{NGC~1566 \textsc{cigale} derived properties}
    \begin{center}
    \begin{tabular}{c|c c c}
    \hline
    \multicolumn{4}{c}{\textit{NUV}-band \textsc{cigale} properties}\\
 \hline
    scale level & Age & Mass & $E(B{-}V)$\\
    parsec  & log$_{10}$(yr) & log$_{10}$(M$_{\odot}$) & \\
      & med  (.25Q,.75Q)& med  (.25Q,.75Q) & med  (.25Q,.75Q)  \\    
    \hline    
    16~pc & 6.6 (6.48,6.78) & 3.62 (3.2,4.02) &  0.15 (0.06,0.25)  \\
    32~pc & 6.7 (6.48,6.85) & 3.78 (3.39,4.21) & 0.15 (0.06,0.26)  \\
    64~pc & 6.85 (6.6,6.9) & 4.04 (3.62,4.48) & 0.14 (0.04,0.25)  \\
    \hline   
    \hline
    \multicolumn{4}{c}{V-band \textsc{cigale} properties}\\
    \hline
    16~pc& 6.78 (6.6,7.85) & 3.75 (3.34,4.11) & 0.16 (0.04,0.28)\\
    32~pc& 6.85 (6.7,7.76) & 3.91 (3.48,4.27) & 0.16 (0.05,0.29) \\
    64~pc& 6.9 (6.78,7.72) & 4.13 (3.72,4.53) & 0.13 (0.05,0.26)\\
    \hline
    \end{tabular}
    \label{TAB:ncg1566_age_mass}
    \end{center}
    \vspace{-10pt}
    \begin{tablenotes}
    \small
    \item Note: \textsc{cigale} derived properties for both the \textit{NUV} and \textit{V}-band selected associations in NGC~1566: [col2] median ages. [col3] median masses. [col4] median attenuation.
    \end{tablenotes}
\end{table}

\subsection{Dependence on absolute magnitude limit of tracer stars}
%--------------------------------
\label{SEC:NGC3351_Mag_cutoff}

The multiple physical scale levels of the hierarchy already facilitate the comparison between galaxies at different distances. However, the signal-to-noise cuts placed on the tracer stars resulted in different absolute magnitude limits for NGC~1566 and NGC~3351.  We compare the results of selection when a brighter absolute magnitude limit is imposed on the tracer stars for NGC~3351, to enable a fairer comparison to the samples of associations identified in NGC~1566, as the galaxy is more distant and the photometric limit is correspondingly brighter. We perform the watershed analysis for NGC~3351 with brighter limits of $M_\mathrm{V} < -4.2$ and $M_\mathrm{NUV} < -6.3$. We find that the number of detected associations ${\sim}2.4$~times less with brighter magnitude limits. While the average size and age of the associations remains approximately the same with the magnitude limits applied, the average mass increases slightly at all scale levels. The average $\log(M/M_{\odot})$ increases slightly by ${\sim}0.25$~dex at 16~pc, ${\sim}0.20$~dex at 32~pc and ${\sim}0.15$~dex at 64~pc for both the NUV and \textit{V}-band selections. This means that the magnitude limit is systematically cutting small or faint associations that therefore have lower mass even though the size and age distribution remain relatively unchanged. Even with this increase in mass taken into account, NGC~1566 stellar associations still have significantly higher median masses.

%--------------------------------
\subsection{Comparison for Selection of Young Stellar Associations}
\label{SEC:nuv_v}
%--------------------------------

%Comparison between NUV and \textit{V}-band selection
We have tested our method using both \textit{NUV} and \textit{V}-band to select tracer stars.  The \textit{V}-band selection allows for a fair comparison to the compact cluster catalogues which use the \textit{V}-band for cluster detection, while the \textit{NUV}-band selection is done to ensure the inclusion of young associations. The \textit{V}-band tracer stars can be both young and old which allows us to trace stellar associations with a wider range of ages, as shown in Section~\ref{SEC:ages_masses}. However, it also allows for the possibility of associations defined by chance super-positions of tracer stars from different generations of star formation. The \textit{NUV}-band provides a cleaner selection of young tracer stars and a coeval selection for associations since the emission of bright O and B~stars peaks in the UV and have ages of only a few Myr.  We now directly compare the overlap between young ($<10$~Myr) associations found by both bands.

The differences between the young NUV and \textit{V}-band selection is apparent across all smoothing scales.  For NGC~3351, there are 294 \textit{NUV}-band associations with estimated ages less than 10~Myr and 403 \textit{V}-band associations at the smallest smoothing scale of 8~pc. $75$~per~cent of the young \textit{NUV}-band \mbox{8-pc} associations have overlap with the young \textit{V}-band associations. For the \textit{V}-band selection, $55$~per~cent of the young associations overlap with the \textit{NUV}-band associations.  The amount of associations with overlap between the NUV and \textit{V}-band detection only increases slightly with the larger scale levels. At the largest scale level of 64~pc, the percentage of \textit{NUV}-band regions that are also detected in \textit{V}-band increases to $85$~per~cent while the percentage of \textit{V}-band regions that overlap with \textit{NUV}-band detections increases to $66$~per~cent.  While there are associations with estimated ages $<10$~Myr in both \textit{NUV}-band and \textit{V}-band that are unique, the \textit{V}-band detects more associations overall and more young associations with $85{-}90$~per~cent of all regions are detected in the \textit{V}-band. 

Of the young \textit{NUV}-band associations, $2{-}8$~per~cent are contained within larger and older \textit{V}-band associations, with the overlap increasing at larger scale levels. For young \textit{V}-band associations, $0.02{-}5$~per~cent are contained within an older \textit{NUV}-band association. This means that the determined ages of overlapping associations tend to be in agreement regardless of the selection used. 

One possibility for the detection of unique associations in the \textit{V}-band is extinction. If the association lies in a dusty region of the galaxy, it may still be visible in the \textit{V}-band while it is obscured to below the detection threshold in the \textit{NUV}-band. The unique \textit{V}-band regions have on average a slightly higher attenuation with $E(B{-}V) = 0.28$~mag versus the average of $E(B{-}V) = 0.195$~mag for the associations that overlap with NUV and $E(B{-}V) = 0.145$~mag for the unique \textit{NUV}-band associations. 

Another possibility for the larger number of detected young associations with \textit{V}-band is that these could be the result of chance super-positions and not a young association of coeval stars. To test this, we looked at the distribution of both the unique and common associations on a colour--colour diagram. As has be seen in the previous section and in Figure~\ref{FIG:NGC3351_cc_cigMass}, the \textit{V}-band selection has more associations that lie off to the right of the SSP track. Associations that contain a mix of stellar population ages will scatter off to the right of the SSP track. While most of the \textit{V}-band association still lie near the SSP track, the majority of the associations with large scatter in the \textit{V}-band are unique to the \textit{V}-band.

Therefore, the \textit{NUV}-band can provide a cleaner selection of young coeval stellar associations however the detection of these associations is limited by attenuation. The \textit{V}-band selection will allow for detection of associations with a wider ranges of ages and is less effected by dust. Therefore, both selections are useful depending on the science goal. We will use and compare the results from both selections in future papers.

%--------------------------------
\begin{table}
    \caption{NGC~3351 Comparison of young ($<10$~Myr) associations}
    \begin{center}
    \begin{tabular}{c|c c c c}
 \hline
    scale level & nRegions & nOverlap & Region overlap & Region Overlap\\
    parsec  &  &  &  \%-Band & \%-Total \\
    \hline 
8~pc NUV & 294 &	221 & 75\%  & 46\%  \\
8~pc \textit{V}-band	& 403 & 221 & 55\% &  46\%  \\

16~pc NUV	& 704  & 586  & 83\%  &  53\%  \\
16~pc \textit{V}-band	& 986 & 587 & 60\% & 53\%  \\

32~pc NUV	& 716  & 590  & 82\%  &  53\%  \\
32~pc \textit{V}-band	& 987  & 592 & 60\% & 53\%  \\

64~pc NUV	& 500  & 427  & 85\%  &  59\%  \\
64~pc \textit{V}-band	& 642  & 425 & 66\% & 59\%  \\
    \hline    
    \end{tabular}
    %\label{TAB:ncg3351_age_mass}
    \end{center}
    \vspace{-10pt}
    \begin{tablenotes}
    \small
    \item Note: The comparison of overlap between \textit{NUV} and \textit{V}-band selected young ($<10$~Myr) associations: [col2] number of regions, [col3] number of regions that have overlap with the other selection band, [col4] percent of the regions in that selection band that overlap with the other selection, [col5] percent of all \textit{NUV} and \textit{V}-band regions that overlap in that selection band. 
    \end{tablenotes}

\end{table}
%--------------------------------

%--------------------------------
\section{RESULTS: Comparison to other data sets}

%--------------------------------
\label{SEC:results2}
%--------------------------------
\subsection{Comparison to LEGUS Star Clusters and Associations}
%--------------------------------

We compare the previously found LEGUS Class~1, 2, and~3 objects in our two test galaxies to our watershed-defined stellar associations.
LEGUS Class~1 and~2 are compact single-peaked clusters while LEGUS Class~3 are multi-peaked objects with asymmetric profiles. The LEGUS Class~3 are most likely compact associations of stars with projected sizes of a few 10s of parsecs and are bright concentrations of star formation in larger stellar associations \citep{adamo17}.
Our regions from the \mbox{16-pc} scale level are therefore closest in size to the LEGUS definition of Class~3 while our larger scale levels of 32 and 64~pc trace the larger structure of the surrounding stellar associations. Since LEGUS uses the \textit{V}-band for cluster identification, our \textit{V}-band selected associations provide the closest comparison to their cluster studies.

For NGC~3351 at the \mbox{16-pc} smoothing level, we find 1401 \textit{V}-band associations while LEGUS found a total of 118 Class~1, 80 Class~2, and 94 Class~3 objects. 
For NGC~1566 at the \mbox{16-pc} scale level, we find 2230 \textit{V}-band associations while LEGUS found a total of
307 Class~1, 271 Class~2, and 469 Class~3 objects. The LEGUS cluster classification is well defined and optimised to find the compact Class~1 and~2 clusters using concentration indices. The LEGUS Class~3 objects are identified by visual inspection of the cluster candidates as regions that are multi-peaked and not compact. The watershed technique adopted in this paper is designed to find regions of multi-peaked star formation at distinct scale levels. If all \textit{V}-band associations at the \mbox{16-pc} scale level are considered compact associations, we find $15$~times the number of LEGUS Class~3 for NGC~3351 and $4.8$~times the number of LEGUS Class~3 for NGC~1566 when using the watershed technique. 

We find that the majority of LEGUS-defined Class~3 objects are contained within our associations starting at the \mbox{16-pc} scale level. The statistics for the number and percentage of Class~1, 2, and~3 LEGUS objects that are contained within both the \textit{NUV} and \textit{V}-band traced associations are outlined in Table~\ref{TAB:ngc3351_cluster_stats} for NGC~3351 and Table~\ref{TAB:ngc1566_cluster_stats} for NGC~1566. While the watershed technique is not optimised to identify compact clusters, Class~1 and~2 compact clusters may still lie within larger associations. If the compact clusters get broken into multiple tracer stars by \textsc{dolphot}, they may be identified at the 8 or \mbox{16-pc} scale level. However, if the compact clusters are identified by only a single \textsc{dolphot} source, they would not be included unless they are part of a larger association.

%--------------------------------
%cluster properties
\begin{table}
\begin{center}
\caption{NGC~3351}
\label{TAB:ngc3351_cluster_stats}
\begin{tabular}{ c | c c c c c}
\multicolumn{6}{c}{NGC~3351: \textit{NUV}-band regions compared to LEGUS objects} \\
 \hline
scale level &  &  & Class~1 & Class~2 & Class~3  \\
parsec & pix & arcsec &  number~(\%) & number~(\%) & number~(\%) \\
 \hline

8.0 & 4.16 & 0.165 &  8 , 6.78\% & 22 , 27.5\% & 33, 35.1\%  \\
16.0 & 8.33 & 0.33 &  23, 19.5\% & 56, 70\% & 78, 83\%\\
32.0 & 16.7 & 0.66 &  35, 29.7\% & 64, 80\% & 87, 92.6\%\\
64.0 & 33.3 & 1.32 &  45, 38.1\% & 68, 85\% & 90, 95.7\%\\
\hline
\multicolumn{6}{c}{\textit{V}-band regions compared to LEGUS objects} \\
\hline

8.0 & 4.16 & 0.165 &  17, 14.4\%  & 29, 36.2\%   & 41, 43.6\%  \\
16.0 & 8.33 & 0.33 &  45, 38.1\% & 63, 78.8\% & 87, 92.6 \%\\
32.0 & 16.7 & 0.66 &  61, 51.7\% & 72, 90\%   & 91, 96.8 \%\\
64.0 & 33.3 & 1.32 &  65, 55.1\% & 75, 93.8\% & 93, 98.9\%\\
\hline
\end{tabular}
\end{center}
\begin{tablenotes}
\small
\item Note: The number and percentage of LEGUS Class~1 (column~4), Class~2 (column~5) and Class~3 (column~6) objects contained in \textit{NUV} or \textit{V}-band selected associations for NGC~3351.
\end{tablenotes}
\end{table}

%--------------------------------

%--------------------------------
%cluster properties
\begin{table}
\begin{center}
\caption{NGC~1566}
\label{TAB:ngc1566_cluster_stats}
\begin{tabular}{ c | c c c c c}
\multicolumn{6}{c}{NGC~1566: \textit{NUV}-band regions compared to LEGUS clusers} \\
 \hline
scale level &  &  & Class~1 & Class~2 & Class~3  \\
parsec & pix & arcsec &  number~(\%) & number~(\%) & number~(\%) \\

 \hline
16.0 & 4.63 & 0.183 & 21, 6.84\% & 86, 31.7\% & 234, 49.9\%\\
32.0 & 9.26 & 0.367 & 59, 19.2\% & 167, 61.6\%& 381, 81.2\%\\
64.0 & 18.5 & 0.733 & 78, 25.4\% & 181, 66.8\% & 390, 83.2\%\\

\hline
\multicolumn{6}{c}{\textit{V}-band regions compared to LEGUS objects} \\
\hline

16.0 & 4.63 & 0.183 & 86, 28\%  & 150, 55.4\%  & 320, 68.2\%\\
32.0 & 9.26 & 0.367 & 160, 52.1\% & 234, 86.3\% & 442, 94.2\%\\
64.0 & 18.5 & 0.733 & 179, 58.3\%   & 229, 84.5\% & 442, 94.2\%\\
\hline
\end{tabular}
\end{center}
\begin{tablenotes}
\small
\item Note: The number and percentage of LEGUS Class~1 (column~4), Class~2 (column~5), and Class~3 (column~6) objects contained in \textit{NUV} or \textit{V}-band selected associations for NGC~1566.
\end{tablenotes}
\end{table}
%--------------------------------

\begin{figure}
    \includegraphics[width=0.44\textwidth]{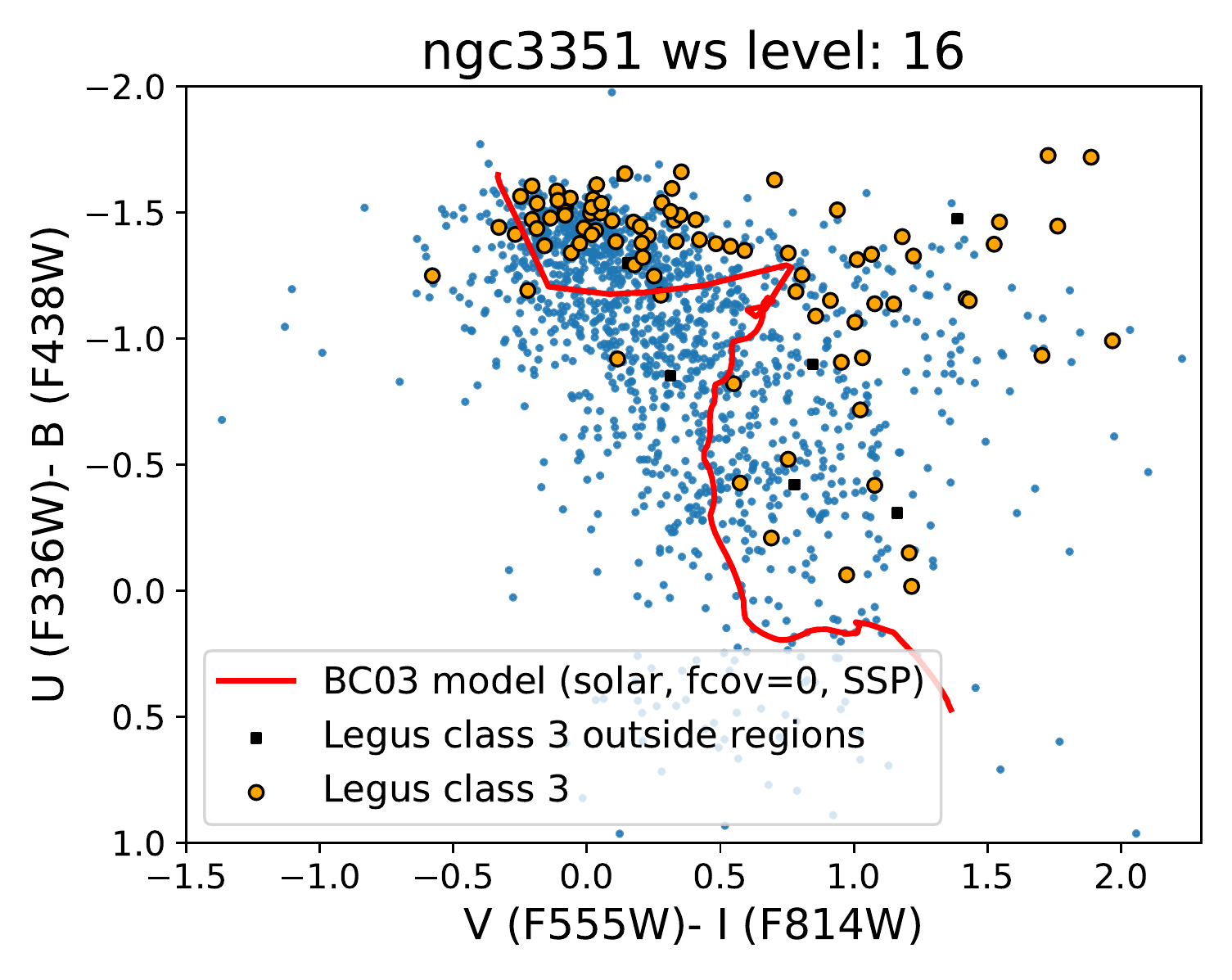}
    \includegraphics[width=0.44\textwidth]{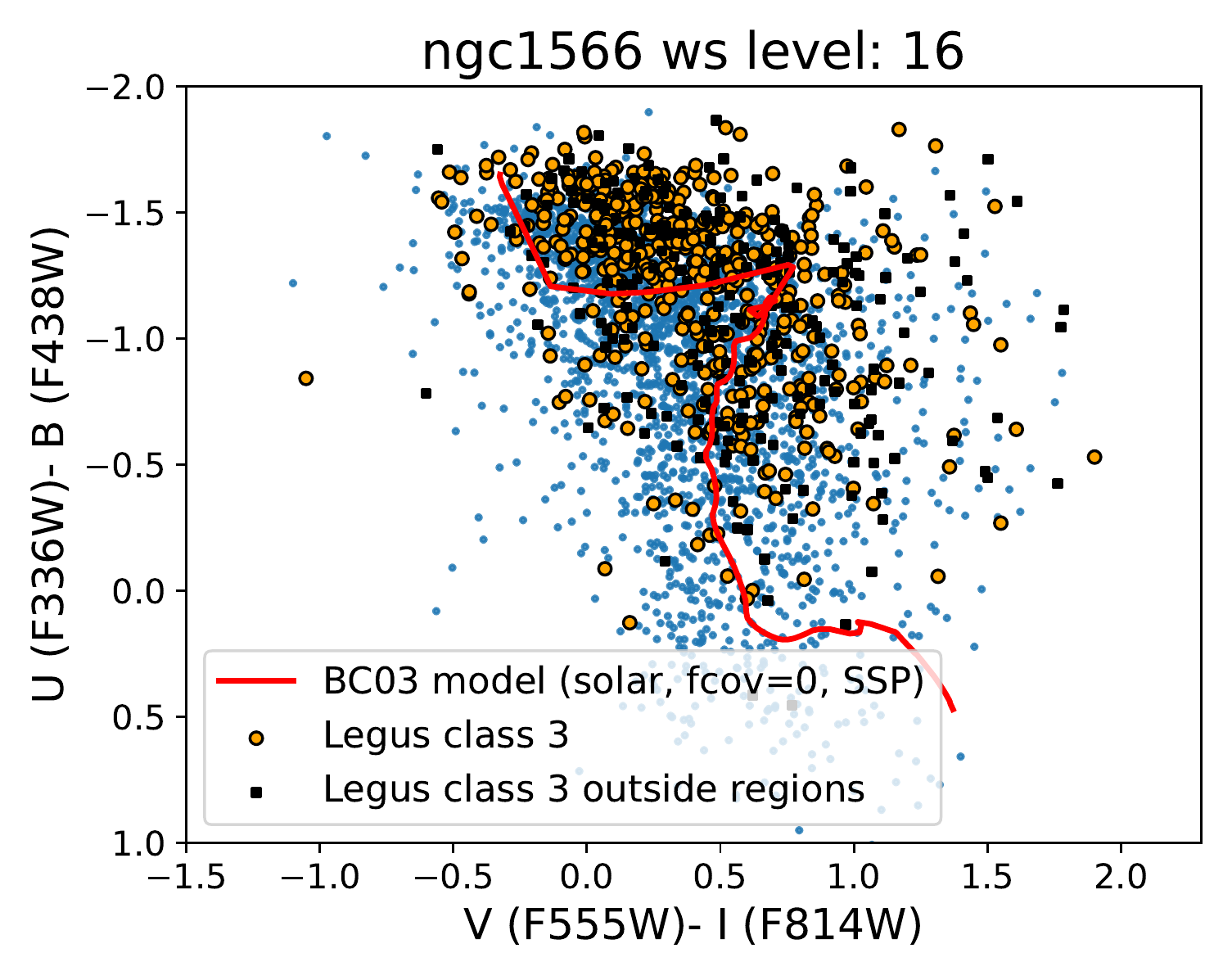}
    \caption{ $UBVI$ colour–colour diagrams for the $V$-band stellar associations (in blue) as well as the LEGUS Class 3 compact associations at the \mbox{16-pc} scale level. The majority of the LEGUS Class~3 compact associations lie within our defined \textit{V}-band regions (shown in orange circles) for NGC~3351 \textit{(top)} and NGC~1566 \textit{(bottom)}. The LEGUS Class~3 compact associations that lie outside of the defined regions (black squares) tend to be older or scattered redward of the BC03 SSP track.}
    \label{FIG:Legus3_inout}
\end{figure}

The watershed technique is designed to find stellar associations. For NGC~3351, the \mbox{16-pc} \textit{V}-band associations contain only 36~per~cent of the Class~1 compact clusters but include 93~per~cent of the LEGUS Class~3 compact associations. The 32 and \mbox{64-pc} smoothing levels contain respectively 97~per~cent and 99~per~cent of the Class~3 compact associations. For NGC~1566, the \mbox{16-pc} \textit{V}-band associations contain 28~per~cent of the Class~1 clusters and 68~per~cent of the LEGUS Class~3 compact associations. The percentage of LEGUS Class~3 compact associations increases to 95~per~cent and 94~per~cent at the 32 and \mbox{64-pc} smoothing levels.  The watershed-defined associations are finding the majority of the LEGUS Class~3 compact associations. Figure~\ref{FIG:Legus3_inout} shows the UBVI colour--colour diagrams for the \textit{V}-band stellar associations (in blue) as well as the LEGUS Class~3 compact associations at the \mbox{16-pc} scale level. The LEGUS Class~3 compact associations that are found within the watershed stellar associations are shown in orange while the LEGUS Class~3 objects not contained within the associations are black squares. 

%---------------------------------
%--------------------------------

%--------------------------------

\begin{figure}
    
    \includegraphics[width=0.48\textwidth]{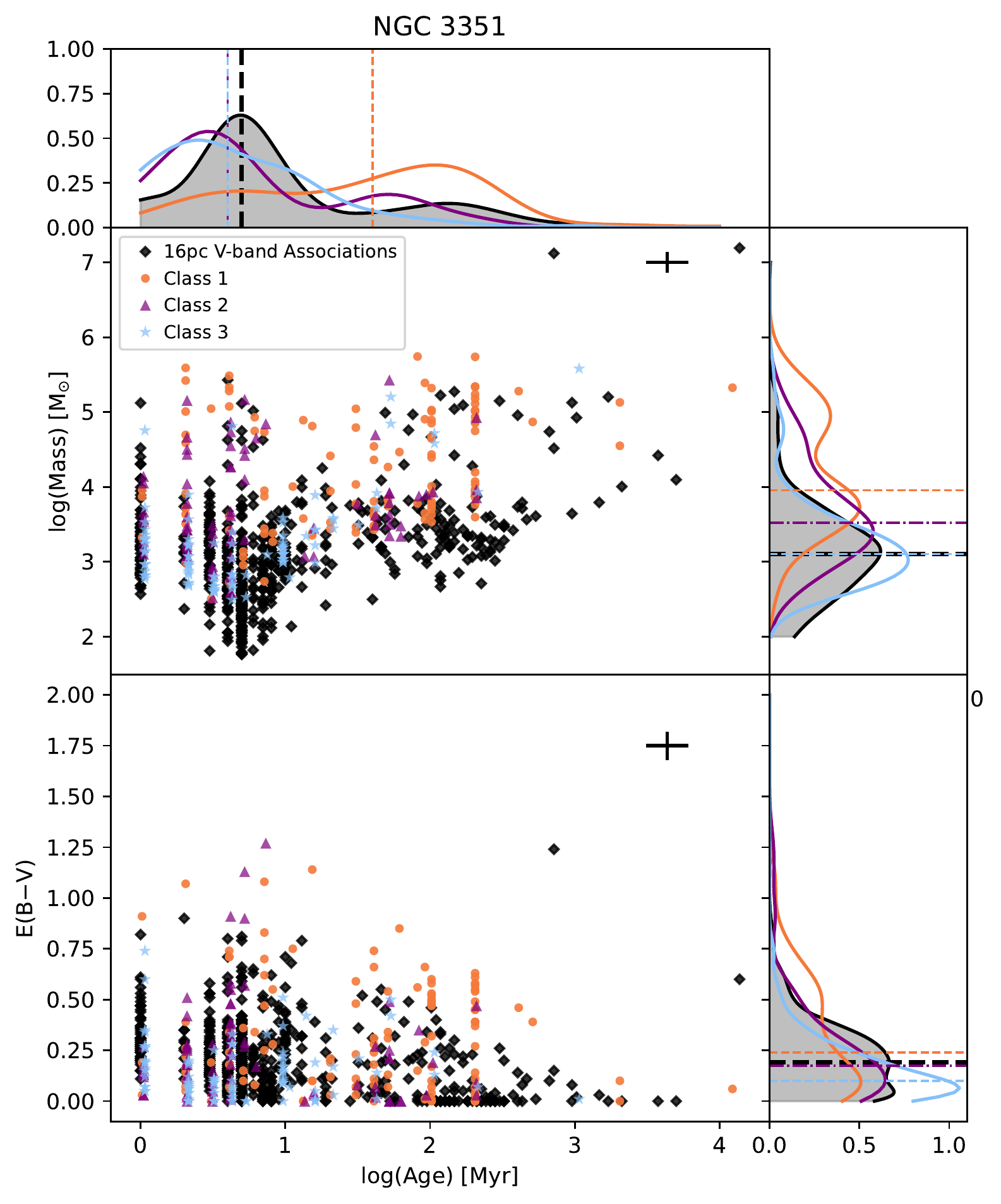}

    \includegraphics[width=0.48\textwidth]{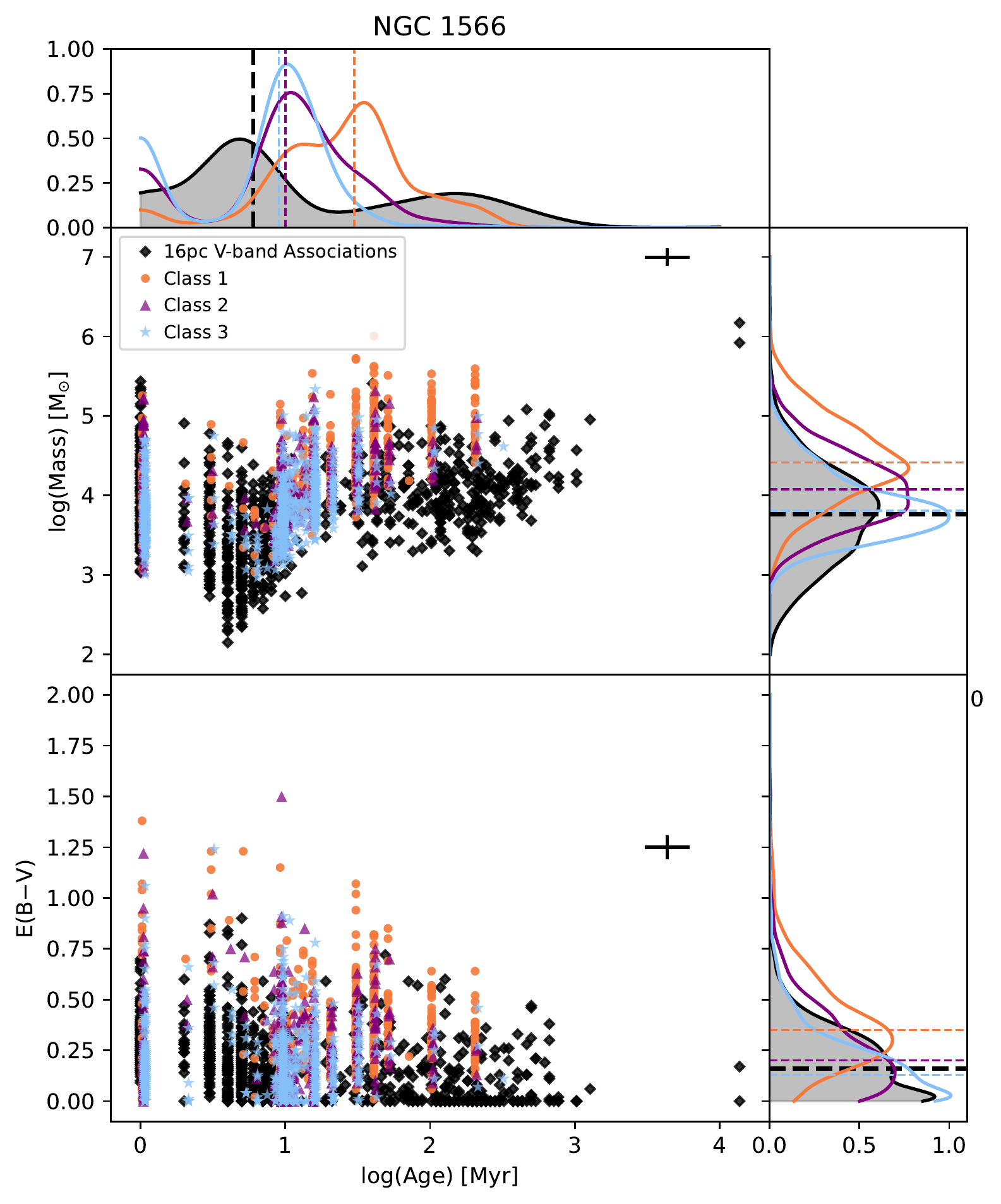}
    
    \caption{Comparison for NGC~3351 \textit{(top)} and NGC~1566 \textit{(bottom)} of the \mbox{16-pc} \textit{V}-band watershed associations (black diamonds) chi-squared estimated ages, masses, and $E(B{-}V)$ to LEGUS objects Class~1 (blue stars), Class~2 (purple triangles), and Class~3 (orange circles). The top histogram in each figure shows the distributions of ages for the \mbox{16-pc} watershed associations and LEGUS objects while the side histograms shows the distribution of the masses and $E(B{-}V)$.}
    \label{FIG:Age_mass_clusters}
\end{figure}

Figure~\ref{FIG:Age_mass_clusters} compares the chi-squared estimated ages, masses, and extinctions of \mbox{16-pc} \textit{V}-band associations for NGC~3351 (top) and NGC~1566 (bottom) to the LEGUS Class~1, 2, and~3 objects. For NGC~1566, the median masses for the LEGUS objects are less than that of the \mbox{16-pc} regions. The lower masses for the LEGUS clusters could be due to their compact sizes. 
For NGC~3351, the median ages of both the Class~2 and~3 objects are similar to the median ages of the watershed associations while the Class~1 compact clusters have an older median log(age/yr) of ${\sim}1.5$~dex. 
For NGC~1566, the median ages of all LEGUS objects are higher than the \textit{V}-band watershed associations. The LEGUS Class~3 compact associations are also selecting an older population than the watershed associations which could be why there is, on average, less overlap for the LEGUS objects and watershed associations for NGC~1566 than there is for NGC~3351.

%---------------------------------
%--------------------------------

%--------------------------------
\subsection{Comparison to \texorpdfstring{\textsc{H\,ii}}{HII} Region Population}
%--------------------------------

We compare the spacial distribution of stellar associations to H$\alpha$ images from the PHANGS--H$\alpha$ survey (Razza et al.\ in prep.) as an independent check on our association definition and age estimates. Our watershed process was designed to find young stellar associations. By checking against the distribution of H$\alpha$ emission, we can determine if our algorithm is indeed recovering young stars that are physically associated and thus properly age-dated.
  
The \textit{V}-band tracer stars select associations with a wider range of ages than the \textit{NUV}-band (see Section~\ref{SEC:results}).  We therefore compare the H$\alpha$ emission to age-binned stellar association regions created using the \textit{V}-band tracer stars to see the distribution of both young and old associations.

\begin{figure*}
    \includegraphics[width=0.8\textwidth]{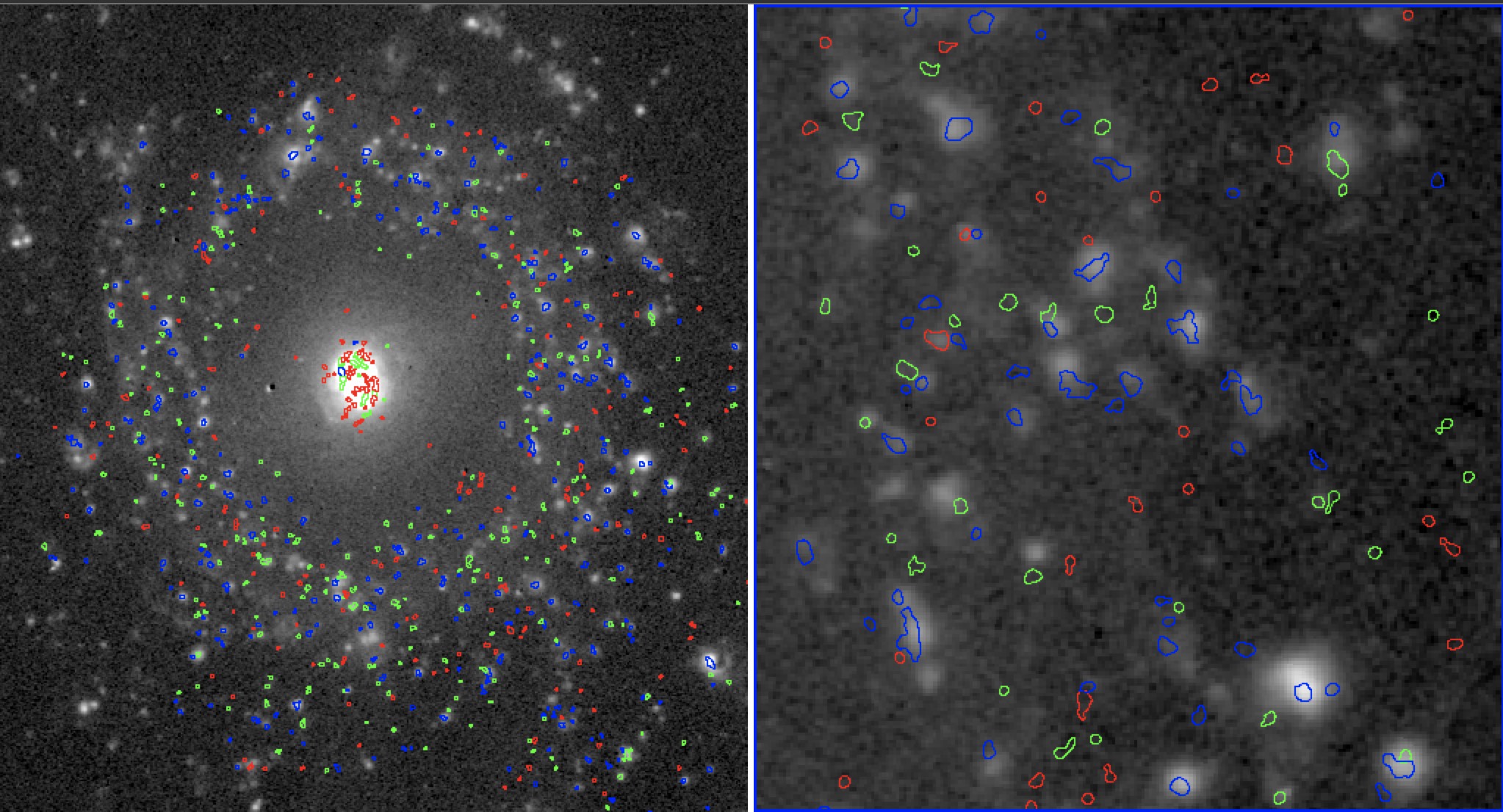}
    \caption{H$\alpha$ image for NGC~3351. Overplotted are the \mbox{32-pc} regions created using the \textit{V}-band tracer stars in age bins ($1{-}3$~Myr in blue, $3{-}5$~Myr in green, and ${>}60$~Myr in red).}
    \label{FIG:NGC3351_halpha}
\end{figure*}

\begin{figure*}
    \includegraphics[width=0.8\textwidth]{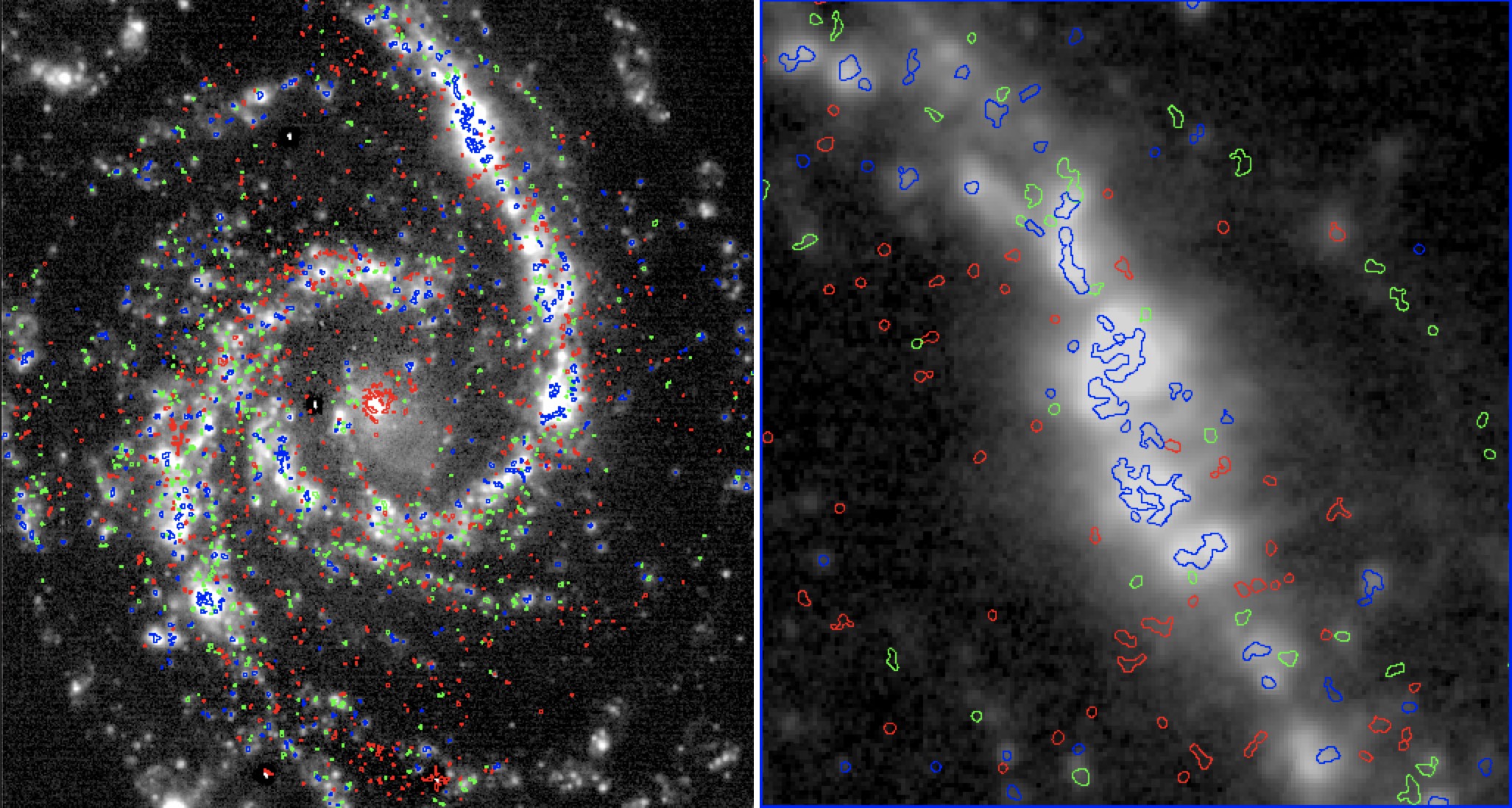}
    \caption{H$\alpha$ image for NGC~1566. Overplotted are the \mbox{32-pc} regions created using the \textit{V}-band tracer stars in age bins ($1{-}3$~Myr in blue, $3{-}5$~Myr in green, and ${>}60$~Myr in red).}
    \label{FIG:NGC1566_halpha}
\end{figure*}

Figures~\ref{FIG:NGC3351_halpha} and~\ref{FIG:NGC1566_halpha} show that the youngest stellar associations with ages of $1$ to $3$~Myr (shown as blue regions) trace the H$\alpha$ emission in both NGC~3351 and NGC~1566.  Stellar associations with slightly older ages of $3$ to $5$~Myr (shown in green) are already less tightly associated with the H$\alpha$ emission.  The oldest stellar associations traced by the \textit{V}-band light with ages greater than $60$~Myr (shown as the red contours) are completely uncorrelated with the brightest H$\alpha$ light.  Since only the youngest ($1{-}3$~Myr) associations trace the H$\alpha$ light, which is only produced by young stars, this is a good independent check that our stellar associations are indeed tracing the young starlight. Future papers by the PHANGS team will perform a detailed analysis of the correlation between stellar associations and a catalogue of H$\alpha$-identified \textsc{H\,ii} regions recently developed \citep[][]{kreckel19,santoro22}.

%---------------------------------
%--------------------------------

%---------------------------------
%--------------------------------

%--------------------------------
\subsection{Comparison to Molecular Cloud Population}
%--------------------------------
\label{SEC:alma}

Using the stellar association catalogues described above, we take an initial look at their correlation with molecular clouds detected in the PHANGS--ALMA \mbox{CO(2--1)} data.  While such analysis will be extensively dev
eloped in forthcoming papers, as it is central to main science goals of PHANGS to map the evolution of multi-scale structure to deduce star formation timescales and efficiencies and investigate molecular cloud life cycle \citep{pan22,kim22,kim21,chevance20,schinnerer19,kruijssen19,kruijssen18,grasha18}, we provide and initial analysis here.  
We examine the extent of pixel overlap between PHANGS--ALMA \mbox{CO(2--1)} molecular cloud catalogues \citep[][Hughes et al.\ in prep.]{rosolowsky21} and stellar association regions. 
We use the molecular cloud catalogues developed by the PHANGS project using \textsc{pycprops}, an updated version of the \textsc{cprops} algorithm \citep{rosolowsky06}, described in detail in \citep[][]{rosolowsky21}.

\textsc{pycprops} segments the CO emission into molecular clouds using a seeded watershed algorithm. The algorithm first identifies local maxima from the CO data. Maxima must have a brightness that is as least twice as bright as the local `merge level', the emission level that contains at least one neighbour. 
The minimum number of pixels required to be uniquely  associated with the maxima is determined by the beam size of the data and equal to number of pixels included in 25\% of the solid angle of the beam 
($N > 0.25\Omega_{\rm{bm}}/\Omega_{\rm{pix}}$). 
The identified local maxima are then used to seed the watershed algorithm which segments the CO emission into individual clouds.

Since the \textit{HST} data have no velocity information, we use a velocity-collapsed \textsc{cprops} map.  When multiple CO clouds are detected along the same line of sight, we compare with the most massive cloud detected along the line of sight.  Full details on the construction of the \textsc{cprops} molecular cloud catalogues and associated maps are given in \citet{rosolowsky21}.

\begin{figure*}
    \includegraphics[width=0.8\textwidth]{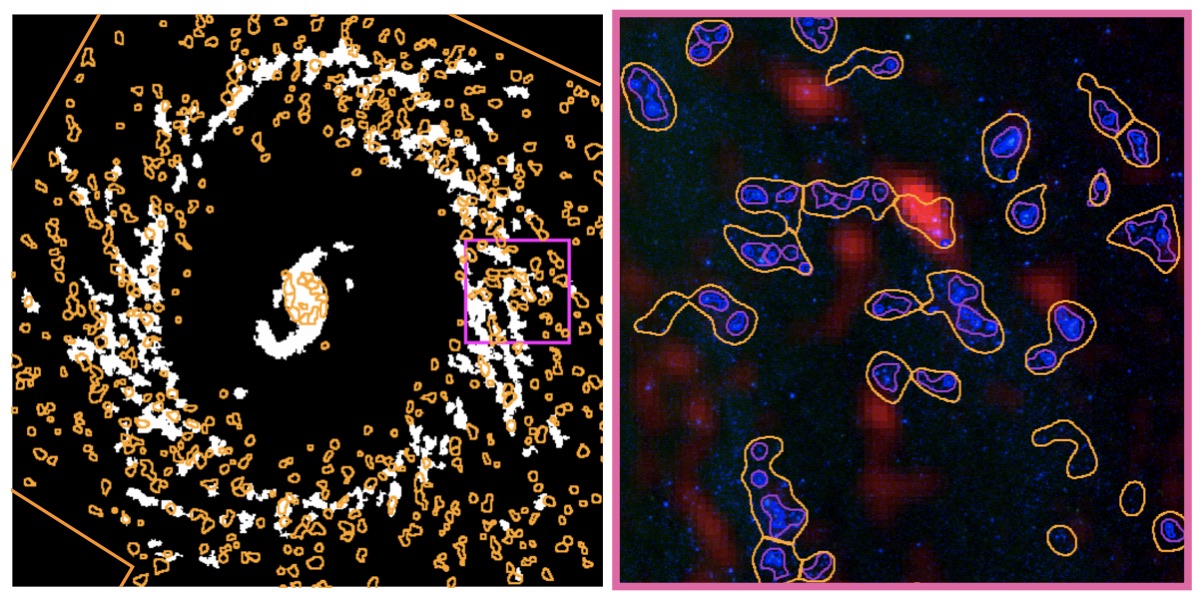}
    \caption{\textit{(Left)} ALMA \textsc{cprops} regions for NGC~3351 in white. Overplotted are the \mbox{64-pc} regions  (orange) created using the NUV-tracer stars and a pink box denoting the region of the zoom in panel. The \textit{HST} FOV is larger than the FOV for the ALMA data and is shown as an orange box. \textit{(Right)} Three colour image of the zoom-in panel with ALMA in red, F275W \textit{HST} in blue and F555W \textit{HST} in green. The corresponding \mbox{64-pc} (orange), \mbox{32-pc} (purple), and \mbox{16-pc} (blue) \textit{NUV}-selected associations are overplotted as regions.}
    \label{FIG:NGC3351_alma}
\end{figure*}

\begin{figure}
    \includegraphics[width=0.48\textwidth]{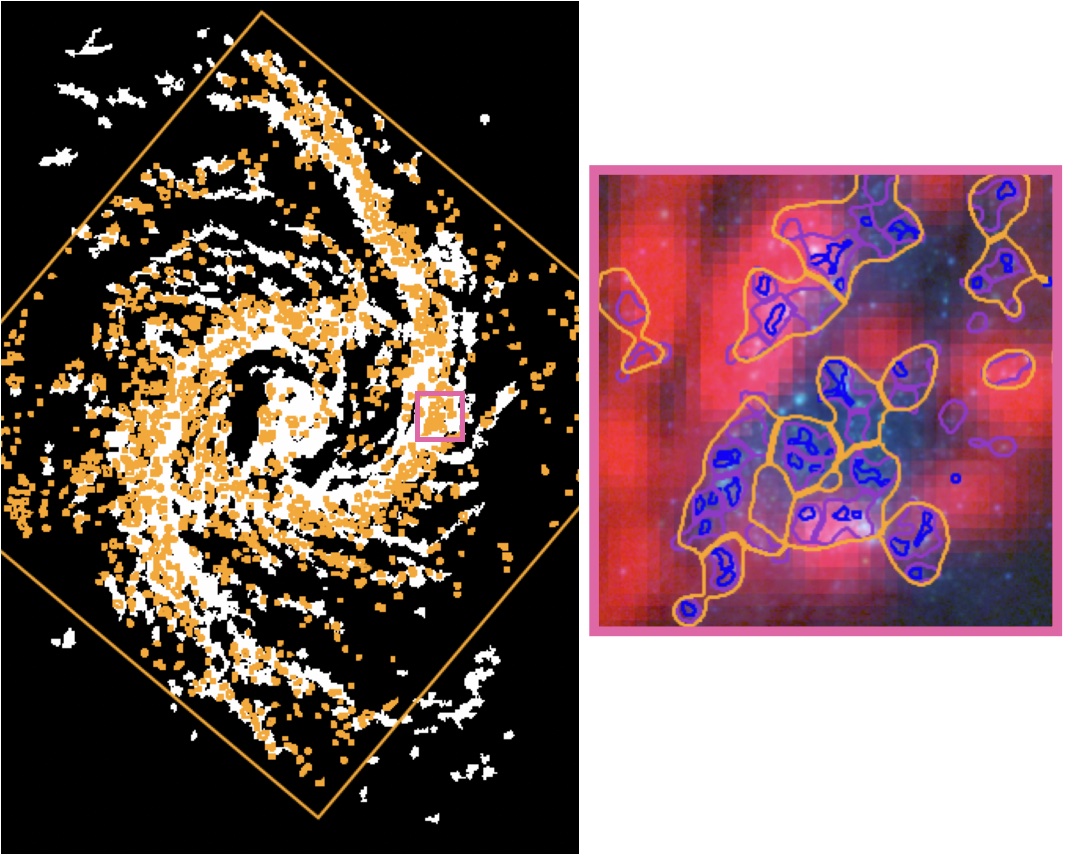}
    \caption{\textit{(Left)} ALMA \textsc{cprops} regions for NGC~1566. Overplotted are the \mbox{64-pc} regions (orange) created using the NUV-tracer stars. The outline of the \textit{HST} footprint is shown as an orange box and the pink box denotes the region of the zoom in panel. \textit{(Right)} Three colour image of the zoom-in panel with Alma in red, F275W \textit{HST} in blue and F555W \textit{HST} in green. The corresponding \mbox{64-pc} (orange), \mbox{32-pc} (purple), and \mbox{16-pc} (blue) \textit{NUV}-selected associations are overplotted as regions. }
    \label{FIG:NGC1566_alma}
\end{figure}

Figures~\ref{FIG:NGC3351_alma} and~\ref{FIG:NGC1566_alma} show the velocity-collapsed \textsc{cprops} map, in grey scale, with the \mbox{64-pc} \textit{NUV}-selected scale level stellar associations overplotted in orange. For our initial comparison, we use the \textit{NUV}-selected associations since (i) we are not interested in the older associations found with the \textit{V}-band and (ii) to ensure a more coeval selection of stellar associations.  For NGC~3351, the ALMA field is fully covered by PHANGS--HST.  For NGC~1566, the PHANGS--HST data cover the majority of the ALMA map, but does not fully extend to the edges. For consistency, we will only consider regions with both ALMA and \textit{HST} coverage.

We calculate the percent overlap between the stellar association regions and the velocity-collapsed \textsc{cprops} map. We find the percent area of overlapping \textsc{cprops} and NUV-selected stellar association regions by doing a pixel-by-pixel comparison of the maps. We also calculate the percentage of stellar association regions that overlap with \textsc{cprops} clouds at every stellar association scale level.

The ALMA beam size for NGC~3351 is $71$~pc while the beam for the more distant NGC~1566 is $110$~pc. This means that our \mbox{64-pc} scale level is closest to the beam size of the ALMA data. In the central ring of NGC~3351, the NUV-selected stellar associations trace a star-forming ring that overlaps the \textsc{cprops} regions. The broader NUV light mostly avoids the bright ALMA \textsc{cprops} regions in the flocculent spiral arms of NGC~3351. When comparing the percent pixel overlap between the \mbox{64-pc} scale level and the \textsc{cprops} regions, only $15$~per~cent of the \textsc{cprops} area overlaps the NUV-selected stellar associations while $18$~per~cent of the area covered by the \mbox{64-pc} scale level associations overlaps with the \textsc{cprops} regions as shown in Table~\ref{TAB:cprops}. 

For NGC~1566, the two brightest spiral arms show overlap between NUV-selected star formation and \textsc{cprops} regions. Therefore, the percentage of pixel overlap of the stellar associations is higher than seen for NGC~3351 at every scale level with $55$~per~cent of the \mbox{64-pc} scale level area overlapping with the \textsc{cprops} regions, see Table~\ref{TAB:cprops}. Even though significantly more of the stellar associations in NGC~1566 overlap with \textsc{cprops} clouds than in NGC~3351, the percentage of the \textsc{cprops} area that contains \mbox{64-pc} stellar associations is similarly small at only $12$~per~cent. This could be due in part to the larger ALMA beam size of $110$~pc for NGC~1566.  

%--------------------------------
\begin{table*}
    \caption{\textit{HST} NUV Stellar association overlap with ALMA \textsc{cprops} catalogues}
    \begin{center}
    \begin{tabular}{c|c c c c}
    NGC~3351:\\
 \hline
    scale level & NUV association &  \textsc{cprops}  & NUV association & \textsc{cprops} \\
      pc  &\% pixel overlap  & \% pixel overlap&\% region overlap &  \% region overlap \\
    \hline    
    16 & 37 &	2 &     34 & 33 \\
    32 & 25 &	6 &     29 & 46 \\
    64 & 18 &	15&     29 & 58 \\
    \hline    

    NGC~1566:\\
 \hline
    scale level & NUV association &  \textsc{cprops}  & NUV association & \textsc{cprops} \\
      pc  &\% pixel overlap  & \% pixel overlap&\% region overlap &  \% region overlap \\
    \hline    
    16 & 76 &	1 &     75 & 27 \\
    32 & 64 &	5 &     69 & 48 \\
    64 & 55 &	12  &   67 & 58 \\
    \hline    
    \end{tabular}
    \end{center}
    \label{TAB:cprops}
    \begin{tablenotes}
    \item %NOTES:
    \end{tablenotes}
\end{table*}
%--------------------------------

As the physical scale of the stellar associations increases, the percentage of the total association area coincident with molecular gas decreases, likely because the larger regions include more area that has not participated in recent star formation. Concurrently, the percentage of the total molecular gas overlap increases, but not in a manner that also results in an increase of the total association area that overlaps.

The majority of the CO clouds by number do not have any overlap with the stellar associations with $67$ to $73$~per~cent of the molecular clouds having no corresponding associations at the smallest common scale level of 16~pc.  This decreases to $42$~per~cent for both galaxies at the largest stellar scale level of 64~pc.  This fraction is similar in both galaxies, which is interesting because in NGC~1566 a much larger percentage of the stellar association regions have overlap with \textsc{cprops} regions than in NGC~3351. The percentage of \mbox{64-pc} stellar associations that overlap with the \textsc{cprops} regions $67$~per~cent for NGC~1566 and only $30$~per~cent for NGC~3351. Therefore, more of the stellar associations are correlated with the molecular clouds in NGC~1566 than in NGC~3351.
These fractions can provide a constraint on the timescales over which clouds appear inactive \citep{pan22,kruijssen18,kruijssen14,schruba10}.

In future studies, we will also investigate using the ALMA CO data to help constrain the extinction values to remove degeneracies in the age estimates. However, in the densest molecular clouds it may be impossible to detect ongoing embedded star formation with our UV and optical data. Upcoming infrared data from \textit{JWST} surveys will be imperative to complete our view of the star formation life-cycle from molecular clouds to fully emerged star clusters and associations.

%---------------------------------
%--------------------------------

%---------------------------------
%--------------------------------

%--------------------------------
%--------------------------------
\section{Discussion}
\label{SEC:discussion}

Future work is still needed to enable consistent comparison across the entire galaxy sample.  PHANGS--ALMA has provided data with matched smoothing scales across the entire sample to allow for consistent comparison of the molecular cloud properties. For the \textit{HST} sample, the galaxies at farther distances, like NGC~1566, will have a higher absolute magnitude limit than nearer galaxies, like NGC~3351. We can make an initial approximation how distance affects the consistent detection of associations across the scale levels by testing how the absolute magnitude limit of the tracer stars affects the results. We will look more closely into the potential effects of distance as we seek to do future cross-galaxy correlations.

One advantage to the scale level and watershed hierarchy method we have developed is that it gives consistent physical levels that can be compared directly across a galaxy sample.  This allows us to probe the smallest resolvable structures at all distances while still connecting the smaller scale structures to the larger scales probed by more distant galaxies.  The multi-scale levels also facilitates direct comparison to ancillary data with a variety of resolutions.  Since we are already smoothing the tracer star maps to different physical scale levels for detection, this could potentially eliminate the need to resmooth the initial data to a common resolution and remeasure everything to compare galaxies at different distances like has been necessary for consistency in other studies of both stellar and CO data \citep[][]{hughes13, larson20, rosolowsky21}.   
Future work can be done to check this directly with the larger galaxy sample.

%--------------------------------
\section{Summary and Conclusions}
\label{SEC:conclusions}
%--------------------------------
We present our method of using a watershed algorithm with \textsc{dolphot} tracer stars to trace stellar associations on a range of scales from 8 to 64~pc. We test different parameters for the watershed algorithm including image filtering and using both NUV and \textit{V}-band selected tracer stars. The properties of the resulting associations are characterised including their sizes, number of tracer stars, and SED-fitted ages, masses and extinction. We analyse the extent to which these regions resemble single age stellar populations by comparing their colours to that of BC03 SSP model track. Finally, we perform an initial comparison of the our watershed-defined stellar associations to previously identified LEGUS star clusters, ground-based H$\alpha$ imaging, and PHANGS--ALMA CO cloud catalogues.

\begin{itemize}
    \item We find that our application of the algorithm successfully identifies populations that are well-characterised by single-age stellar populations, and occupy loci in the $UBVI$ colour--colour diagram consistent with previously published catalogues of compact clusters and associations. 
    
    \item  The median log(age/yr) is the same for both NGC~3351 and NGC~1566 and increases at larger smoothing scales from of $\log (\mathrm{age/yr}) = 6.6{-}6.9$.
    The \textit{V}-band tracer stars identify not only young associations but also older associations resulting in a median log(age/yr) that is ${\sim}0.15$~dex older than the \textit{NUV}-band selection at all all scale levels.  
    
    \item The stellar associations range in mass from $\log(M/M_{\odot}) = 3{-}7$. As the associations increase in size at larger scale levels, the median mass also increases by ${\sim}1.6$~times from the 16 to \mbox{64-pc} scale level for both galaxies. The masses of the stellar associations identified in NGC~1566 are on average $1.9$~times the more massive than the associations in NGC~3351 at each corresponding scale level. 
    
    \item We compare the agreement in young ($<10$~Myr) associations identified with \textit{NUV} and \textit{V}-band tracer stars in NGC~3351. We find that there is $46{-}59$~per~cent overlap in the corresponding young associations from 8 to 64~pc. Since the \textit{V}-band selection identifies more associations at all scale levels, the majority of the unique associations are from the \textit{V}-band selection where $75{-}85$~per~cent of the young \textit{NUV}-band selected associations overlap with the larger sample of \textit{V}-band selected associations. Two factors seem to be contributing to the unique detections in \textit{V}-band: attenuation and chance super-positions of stars from uncorrelated populations. Therefore, the \textit{NUV}-band selection can provide a cleaner selection of young coeval stellar associations; however, the \textit{V}-band selection is necessary if associations with a wider range of ages are desired. 
    
    \item We compare our the overlap of our regions to previously identified LEGUS objects. Most of the LEGUS Class~3 compact associations are contained within the watershed-defined associations by the \mbox{16-pc} smoothing level with $93$~per~cent for NGC~3351 and $68$~per~cent for NGC~1566 of the LEGUS Class~3 objects being contained within the \textit{V}-band watershed associations. The percentage of included LEGUS objects increases at larger smoothing scales.
    The \mbox{16-pc} \textit{V}-band watershed associations for NGC~3351 have the same median age as LEGUS Class~2 and Class~3 objects while the \mbox{16-pc} \textit{V}-band watershed associations for NGC~1566 are on average younger than all classes of LEGUS objects. 

    \item We compare the spatial distribution of stellar associations to H$\alpha$ emission as an independent check on our association definition and age-fitting. We find that the youngest stellar associations with ages between $1{-}3$~Myr trace the H$\alpha$ distribution of light. By $3{-}5$~Myr, the stellar associations already start to become less associated with the H$\alpha$. Older stellar associations with ages greater than $60$~Myr are completely unassociated with the H$\alpha$ light. 

    \item We compare the overlap between \textit{NUV}-selected associations and velocity-collapsed \textsc{cprops} molecular cloud catalogues. We find that the percentage of \textsc{cprops} regions that have overlap with \mbox{64-pc} stellar associations is ${\sim}58$~per~cent for both galaxies. However, the percentage of \mbox{64-pc} stellar associations that overlap with the \textsc{cprops} regions is only $29$~per~cent for NGC~3351 and $67$~per~cent for NGC~1566. Therefore, more of the stellar associations are correlated with the molecular clouds in NGC~1566 than in NGC~3351.
 
    \item The correlation between the \textsc{cprops} molecular clouds and \textit{NUV} stellar associations varies with environment in the galaxy. In the dense spiral arms of NGC~1566, there are large overlaps between \textsc{cprops} regions and stellar associations. In more diffuse CO regions like the inter-arm region of NGC~1566 and disk of NGC~3351, there is less overlap between the \textsc{cprops} regions and stellar associations.

\end{itemize}

The method presented here will be applied to all 38 galaxies in the PHANGS--HST sample. This paper provides a first-look at the comparisons that can be performed using our multi-scale stellar association catalogues. The stellar associations are vital to studying the stellar life-cycle from molecular clouds to emerging stellar complexes.

%--------------------------------
\section{Acknowledgements}
%--------------------------------
This research is based on observations made with the NASA/ESA Hubble Space Telescope obtained from the Space Telescope Science Institute, which is operated by the Association of Universities for Research in Astronomy, Inc., under NASA contract  NAS 5–26555.  Support for Programme number 15654 was provided through a grant from the STScI under NASA contract NAS5-26555.

This paper makes use of the following ALMA data: \linebreak
ADS/JAO.ALMA\#2015.1.00925.S, \linebreak % (pilot low mass)
ADS/JAO.ALMA\#2015.1.00956.S, \linebreak % (pilot high mass)
ALMA is a partnership of ESO (representing its member states), NSF (USA) and NINS (Japan), together with NRC (Canada), MOST and ASIAA (Taiwan), and KASI (Republic of Korea), in cooperation with the Republic of Chile. The Joint ALMA Observatory is operated by ESO, AUI/NRAO and NAOJ.

This research has made use of the NASA/IPAC Extragalactic Database (NED) which is operated by the Jet Propulsion Laboratory, California Institute of Technology, under contract with NASA. 

JMDK gratefully acknowledges funding from the Deutsche Forschungsgemeinschaft (DFG) in the form of an Emmy Noether Research Group (grant number KR4801/1-1), as well as from the European Research Council (ERC) under the European Union’s Horizon 2020 research and innovation programme via the ERC Starting Grant MUSTANG (grant agreement number 714907). COOL Research DAO is a Decentralised Autonomous Organisation supporting research in astrophysics aimed at uncovering our cosmic origins.

SD is supported by funding from the European Research Council (ERC) under the European Union’s Horizon 2020 research and innovation programme (grant agreement no. 101018897 CosmicExplorer).

KK gratefully acknowledges funding from DFG in the form of an Emmy Noether Research Group (grant number KR4598/2-1, PI Kreckel).

TGW acknowledges funding from ERC under the European Union’s Horizon 2020 research and innovation programme (grant agreement No. 694343).

RSK acknowledges support from ERC via the Synergy Grant ``ECOGAL'' (project ID 855130), from the Heidelberg Cluster of Excellence (EXC 2181 - 390900948) ``STRUCTURES'', funded by the German Excellence Strategy, from DFG in the Collaborative Research Center SFB 881 ``The Milky Way System'' (funding ID 138713538, subprojects A1, B1, B2, and B8), and from the German Ministry for Economic Affairs and Climate Action in project ``MAINN'' (funding ID 50OO2206). RSK also thanks for HPC resources and data storage supported by the Ministry of Science, Research and the Arts of the State of Baden-W\"{u}rttemberg (MWK) and DFG through grant INST 35/1314-1 FUGG and INST 35/1503-1 FUGG, and for computing time from the Leibniz Computing Center (LRZ) in project pr74nu. 

KG is supported by the Australian Research Council through the Discovery Early Career Researcher Award (DECRA) Fellowship DE220100766 funded by the Australian Government. 

%%%%%%%%%%%%%%%%%%%%%%%%%%%%%%%%%%%%%%%%%%%%%%%%%%

\section{Data Availability}
%%%%%%%%%%%%%%%%%%%%%%%%%%%%%%%%%%%%%%%%%%%%%%%%%%

The data underlying this article can be retrieved from the Mikulski Archive for Space Telescopes at \url{https://archive.stsci.edu/hst/search_retrieve.html} under proposal GO-15654. High level science products are provided at \url{https://archive.stsci.edu/hlsp/phangs-hst} including science ready mosaicked imaging associated with HST GO-15654 with digital object identifier \doi{10.17909/t9-r08f-dq31} and all multi-scale association regions and calculated properties developed in this paper.

%%%%%%%%%%%%%%%%%%%% REFERENCES %%%%%%%%%%%%%%%%%%

\bibliographystyle{mnras}   
\bibliography{phangshst}

%%%%%%%%%%%%%%%%%%%%%%%%%%%%%%%%%%%%%%%%%%%%%%%%%%

%--------------------------------------------
\appendix 

%--------------------------------------------
\section{High-pass filtering}
\label{SEC:appendix}

\begin{figure}
 \includegraphics[width=0.24\textwidth]{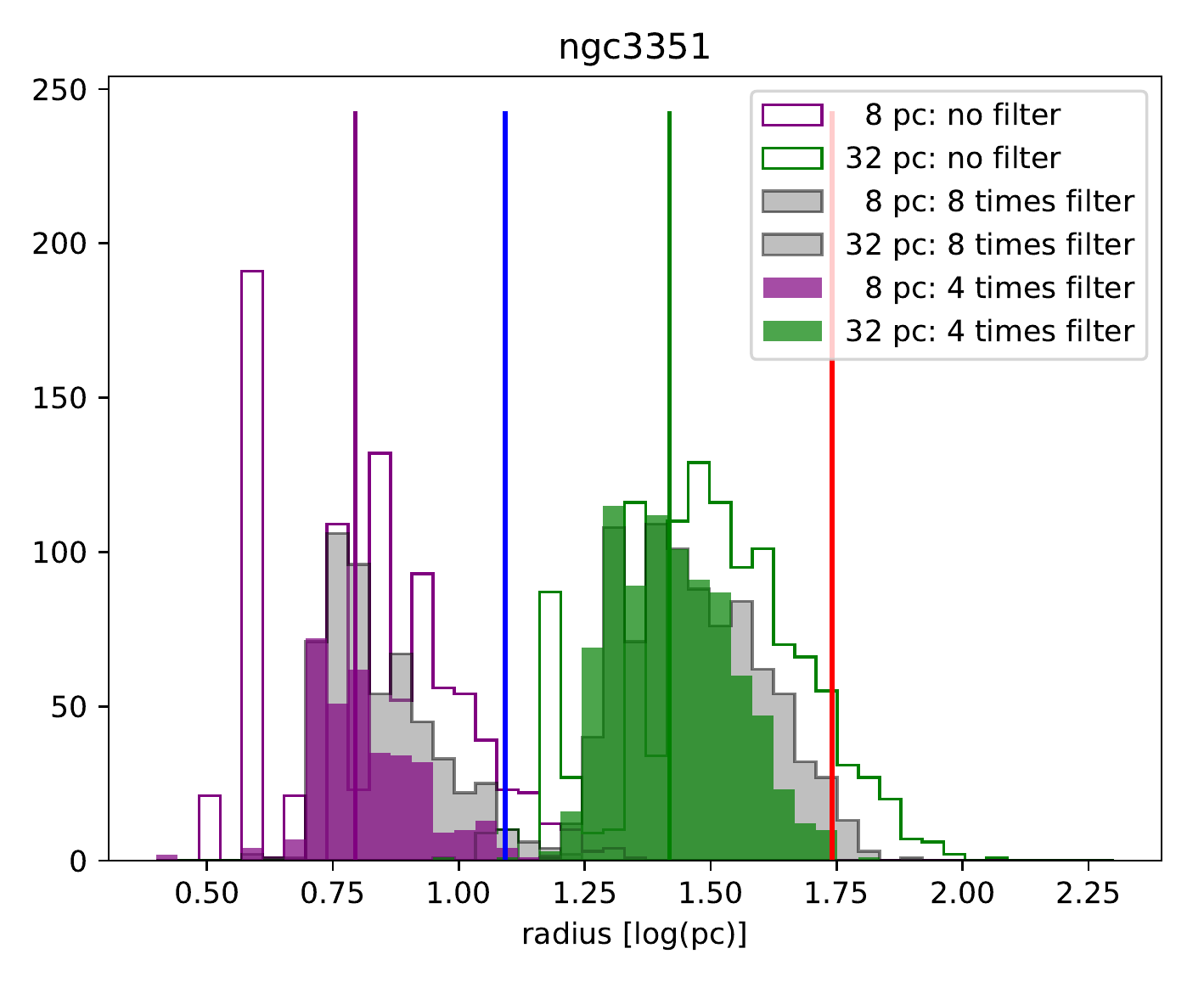}
  \includegraphics[width=0.24\textwidth]{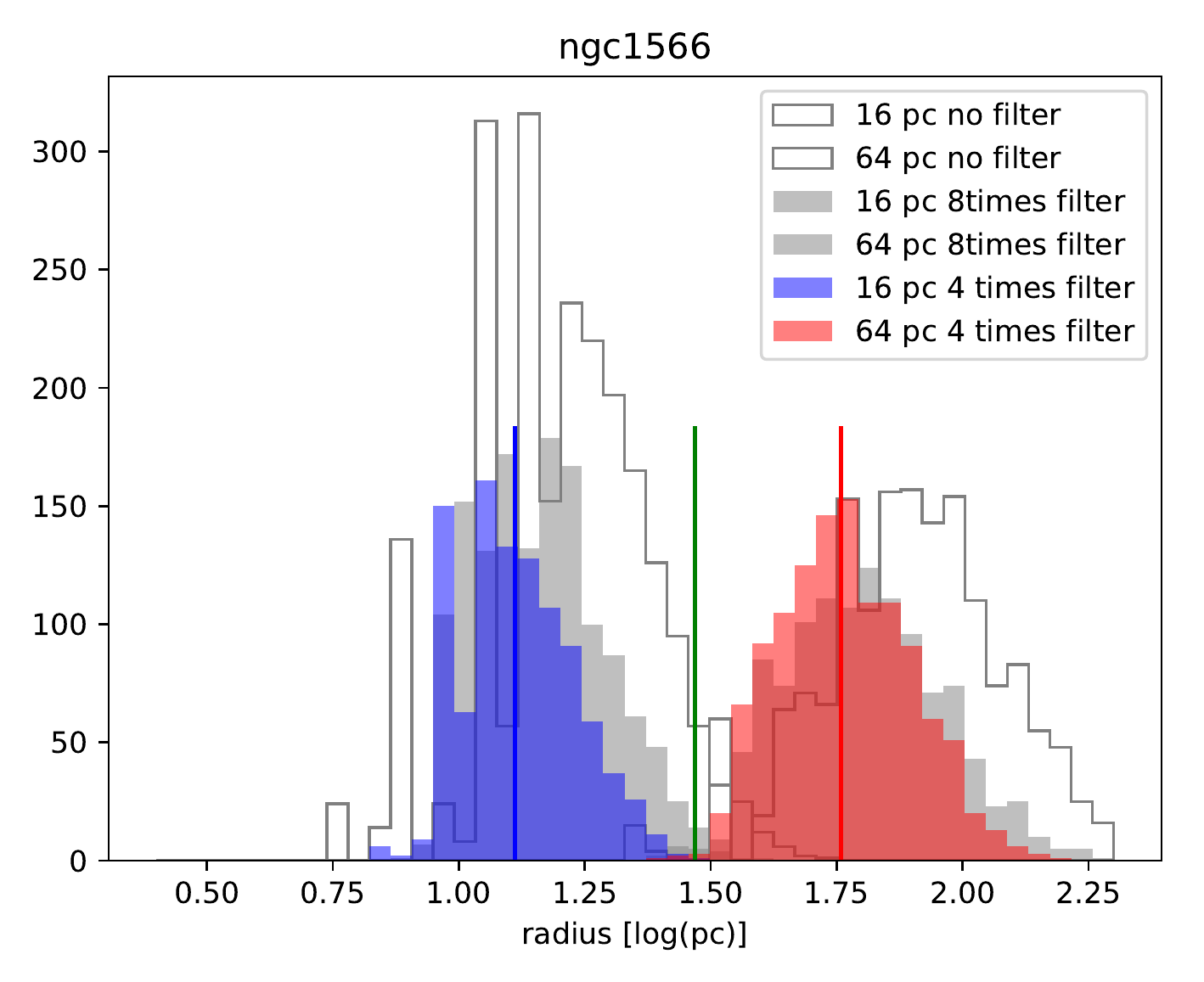}
 \caption{Distribution of region sizes at two smoothing levels for each galaxy. We show how the distribution changes with no filtering applied (empty histogram), and with two different high-pass filtering steps: filter step of 4~times the smoothing level (32~pc to 128~pc) shown as the filled coloured histograms and a filter step of 8~times the smoothing level (32~pc to 256~pc) shown as the grey filled histograms. Vertical lines make the median size values of the 4~times filtered distributions. Both filtering steps give similar median size values for the regions.The 4~times filtering produces the tightest size distribution with the least overlap between scale levels. }
 \label{FIG:SmoothingScales}
\end{figure}

We compare the use of two different high-pass filters.
We subtract a map that is smoothed to a scale 4~times larger than the original map, and another smoothed to scale that is 8~times larger. 
To determine which filtering step may be more appropriate for selection of associations, we analyze the distribution of sizes from the resulting regions compared to those in the original unfiltered image.  
%---- high-pass filtering steps ---
We find that both high-pass filtered images (4~and 8~times larger) have size distributions with median effective radii similar to the FWHM of the Gaussian smoothing kernel (see Table~\ref{TAB:ngc3351_filterstep}).
As expected, the filtering does result in fewer regions detected at each scale level compared to when no filtering is applied. Figure~\ref{FIG:SmoothingScales} shows that the regions that are being filtered out are at the tails of the size distribution and that we are effectively sampling the bulk of the distribution.The high-pass filtering removes more of the larger sized regions at each scale level. This is desired since the scale levels are defined to emphasize different sized associations. The larger structures that are filtered out at the small scale levels will be recovered at larger scale levels. 

%--------------------------------
%filter step  comparison
\begin{table}
    \caption{NGC~3351 filter step comparison.}
    \begin{center}
    \begin{tabular}{ c | c c c c c c c}
    \hline
    scale level  & nRegions& radius \\
    parsec       & & med  (.25Q,.75Q) \\
    \hline
    \multicolumn{8}{c}{NGC~3351, 4~times filter step, NUV tracer stars} \\
    \hline
    8.0 &  334 & 6.23 (5.53,7.66)   \\
    16.0 & 807 & 12.4 (10.3,15.3)   \\
    32.0 & 838 & 26.1 (21.6,32.4)   \\
    64.0 & 641 & 55.2 (45.2,67.4)   \\
    \hline
    \multicolumn{8}{c}{NGC~3351, 8~times filter step, NUV tracer stars} \\
     \hline
    8.0 &  544 & 6.59 (5.53,8.32)  \\
    16.0 & 909 & 13.7 (10.9,17.0)  \\
    32.0 & 883 & 28.4 (23.1,36.5)  \\
    64.0 & 634 & 59.6 (47.3,75.4)  \\
    \hline
    \label{TAB:ngc3351_filterstep}
    \end{tabular}
    \end{center}
    \vspace{-10pt}
    \begin{tablenotes}
    \small
    \item Note: Properties of NUV associations with two different high-pass filters: 4~times filter step and 8~times filter step. [col1] scale level. [col2] number of regions [col3] the median, 1st and 3rd quartile radius of regions. 
    \end{tablenotes}
\end{table}

%2625 total regions
%Number of NUV Tracers Stars selected: 11812
%Number of \textit{V}-band Tracers Stars selected: 20472
%--------------------------------

The high-pass filters effectively remove tails in the distribution at both the high and the low end that are present at all scale levels in the original unfiltered image. Both of the high-pass filtered images therefore have a tighter distribution of sizes than the unfiltered image; however, the larger filter step (8~times larger) produces a wider spread of region sizes than the smaller filter step (4~times larger) and produces size distributions with more overlap between the different scale levels as shown in Figure~\ref{FIG:SmoothingScales}. Since the specific smoothing scales are defined with the purpose of selecting specific sizes of structures and minimizing the overlap between the different scale levels, we adopt the smaller filter step which produces a tighter distribution of region sizes. Therefore, we will use the 4~times high-pass filter as our fiducial choice.

%---------------------------------
%--------------------------------
%--------------------------------

%\clearpage
%\onecolumn

%%%%%%%%%%%%%%%%% APPENDICES %%%%%%%%%%%%%%%%%%%%%

%\appendix

% Don't change these lines
\bsp	% typesetting comment
\label{lastpage}
\end{document}